\RequirePackage{fix-cm}

\documentclass[twocolumn,epjc3]{svjour3}  
\smartqed

\usepackage[T1]{fontenc}
\usepackage{amsmath,amssymb}
\usepackage{graphicx}
\usepackage{hyperref}
\usepackage{cite}
\usepackage{xcolor}
\usepackage{placeins}
\usepackage{listings}

\begin{document}

\title{Sensitivity Analysis of Singlet Vector-Like B Quarks via Photon-Induced and Z-Initiated Processes at FCC–$\mu p$}



\author{Eda Alici$^{*}$}

\institute{
Department of Physics, Zonguldak Bulent Ecevit University,
Zonguldak, Turkey\\
$^{*}$Corresponding author: \href{mailto:eda.alici@beun.edu.tr}{eda.alici@beun.edu.tr}
}
\date{Received: date / Accepted: date}

\maketitle


\begin{abstract}
This study presents a systematic sensitivity analysis of singlet-type vector-like $B$ quark production at an FCC--$\mu p$ collider with a centre-of-mass energy of $\sqrt{s}=24.5~\mathrm{TeV}$ through photon- and $Z$-initiated production mechanisms. The analysis focuses on the $B\to Zb$ decay channel, considering the leptonic decay of the $Z$ boson and the hadronic decay of the accompanying $W$ boson, leading to the final state $\ell^+\ell^-bjj$. Detector-resolution effects are incorporated through a simplified Gaussian smearing procedure, and the discovery and exclusion sensitivities are evaluated using an Asimov-based statistical framework.

For the photon-induced channel, the most favourable sensitivity is obtained for $R_L=0.05$ and $\mathcal{L}=1000~\mathrm{fb^{-1}}$. Over the mass range $M_B=2$--$3~\mathrm{TeV}$, the expected $5\sigma$ discovery reach is approximately $g^\ast\simeq0.263$--$0.328$, while the $95\%$ C.L. exclusion sensitivity extends to $g^\ast\simeq0.162$--$0.197$. On the other hand, the $Z$-initiated channel provides a substantially stronger sensitivity and extends the investigated mass range up to $M_B=4.5~\mathrm{TeV}$. For $R_L=0.05$ and $\mathcal{L}=500~\mathrm{fb^{-1}}$, the $5\sigma$ discovery reach is approximately $g^\ast\simeq0.038$--$0.056$, while the corresponding $95\%$ C.L. exclusion sensitivity reaches $g^\ast\simeq0.021$--$0.032$. 

These results demonstrate that photon- and $Z$-initiated single production at an FCC--$\mu p$ collider provide complementary probes of heavy vector-like $B$ quarks. Moreover, the $Z$-initiated channel offers particularly strong sensitivity to small effective couplings in the multi-TeV mass region beyond the present direct LHC reach.
\end{abstract}


\section{Introduction}
\label{sec:intro}

The Standard Model (SM) successfully explains a large portion of the high-precision particle physics experiments conducted to date. However, the SM fails to provide satisfactory solutions to some fundamental problems. Problems such as the mass hierarchy problem, the absence of dark matter candidates, baryon asymmetry, and the origin of neutrino masses are some of the extremely important shortcomings of the SM. These shortcomings necessitate the search for new physics scenarios beyond the SM. In this context, many new physics scenarios are being intensively investigated both theoretically and experimentally. At this point, Vector-Like Quarks (VLQs) are among the candidates for new fermions with strong theoretical motivation and appear naturally in many extended models. ~\cite{VLQ_review1,VLQ_review2,Okada:2012gy,Alves:2024vlq}.

As it is well known, the left- and right-handed components of VLQs have the same transformation properties under electroweak representations. Therefore, VLQ masses can be directly generated via Dirac mass terms and these masses are not necessarily dependent on the Higgs mechanism. This characteristic contributes to alleviate the hierarchy problem and also facilitates the control of the flavour structure. For these reasons, VLQs stand out as attractive particles in theoretical research. On the other hand, VLQs are classified into types $T$, $B$, $X$ and $Y$ according to their electromagnetic charges and electroweak quantum numbers. Furthermore, these different types of VLQs can appear as singlet or doublet under $SU(2)_L$ gauge symmetry.

A broad phenomenological literature has been reported on VLQs with different electroweak representations, production mechanisms, and decay patterns. General and model-independent aspects of VLQ searches, including their collider phenomenology, coupling structures, and possible finite-width effects, have been discussed in Refs.~\cite{Atre2011SingleVLQ,Chala2018VLQFuture,Angelescu2016VLQHiggs,Carvalho2018VLQ,Barducci2017VLQdiscrimination,Banerjee2024VLQStatus}. For $T$-type VLQs and top partners, both single- and pair-production mechanisms have been explored at hadron colliders through several characteristic decay modes~\cite{Cacciapaglia2010BoundsVLQT,Yang2021VLQTProd,Zhou2020NewDecayVLQ,Liu2017SingleTgamma,Liu2020SingleB,DeSimone2013TopPartner}, while complementary studies have also investigated their production at future $ep$, $e^+e^-$, muon, and muon--proton colliders~\cite{Yang2019SingleVLQ,Qin2022VLQT_CLIC,Lv2022MuonVLQ,Han2025MuonProtonVLQ}. The phenomenology of $B$-type VLQs has received particular attention because of the characteristic $B\to Zb$, $B\to Wt$, and $B\to Hb$ decay modes. At hadron colliders, single-production searches have been studied in the $Zb$ and $Wt$ channels at the LHC, HL-LHC, and future high-energy $pp$ colliders~\cite{Gong2019SingleB,Yang:2024VLQBhadronic,Han2022SingleVLQB,Han2022NuPhB975,Canbay2023SingleVLQ,Han2023EPJC846}, while alternative scalar-mediated signatures have also been investigated~\cite{Bardhan2023MLVLQ}. At lepton--hadron colliders, single production of vector-like $B$ quarks has been studied at the LHeC and future $ep$ machines in several final states, including the $bZ$ channel~\cite{ZengPanZhang2023,Gong2020SingleVLQB,Shang2022VLQB}. Complementary studies at high-energy $e^+e^-$ colliders have considered both single and pair production of singlet $B$ quarks at CLIC~\cite{Han2022VLQCLIC,Han2024PairVLQB,Yang2025CLICPairB}. In addition to the conventional $T$- and $B$-type states, searches for exotic-charge VLQs, including $X_{5/3}$ and $Y_{-4/3}$ states, as well as non-minimal VLQ multiplets, have been investigated at hadron and lepton--hadron colliders~\cite{Chen2017SingleB,Liu2024VLQX,Zhang2024VLQX_FCCeh,Cetinkaya2021SingleVLQ,Banerjee2024TripletVLQ}.

In the present work, we investigate the VLQ-($B$) with electric charge $Q_B=-1/3$ and singlet representation $SU(2)_L$. Here, VLQ-($B$) interacts with SM quarks via mixing.  This mixing allows three main decay channels for the $B$ quark:
\begin{equation}
B \to Wt, \qquad
B\to Zb, \qquad
B \to Hb
\end{equation}

In the heavy mass limit, the branching ratios of these channels are approximately $\mathrm{BR}(B\to Wt) : \mathrm{BR}(B\to Zb) : \mathrm{BR}(B\to Hb) \simeq 2:1:1 $ order.
This universal behaviour can be explained by Goldstone's equivalence theorem. However, the exact ratios depend on the mixing structure and model-specific interactions, and these ratios may vary accordingly ~\cite{He1992Equivalence,He1994Equivalence,VLQ_review2,Gong2019SingleB}.

In the simplified model used in this paper, the effective coupling parameter $g^\ast$ determines the overall strength of the mixing-induced interactions of the VLQ-$B$ quark with SM fermions and electroweak bosons. The parameter $R_L$, on the other hand, controls the relative strength of the first- and third-generation interactions, while the second-generation mixing is set to zero in the adopted model. In this study, the benchmark values $R_L=0.05$ and $R_L=0.01$ are considered in order to investigate the effect of a small first-generation contribution on the production sensitivity. In the heavy-mass limit, these choices remain close to the characteristic singlet-VLQ decay pattern $\mathrm{BR}(B\to Wt):\mathrm{BR}(B\to Zb):\mathrm{BR}(B\to Hb)\simeq2{:}1{:}1$. The photon- and $Z$-initiated production mechanisms considered in this work provide complementary probes of the same singlet VLQ-$B$ framework~\cite{Buchkremer2013VLQUFO,ZengPanZhang2023}.

This study focuses on the $B \to Zb$ decay channel. This choice is motivated by the fact that this channel provides an experimentally clean reconstruction and offers clear advantages in terms of signal-to-background separation. These advantages become even more pronounced when the leptonic decay channel of the $Z$ boson is preferred. This is because the leptonic $Z$ decay ($Z \to \ell^+ \ell^-$) allows for clear observation of heavy resonances due to its low SM background and high invariant mass resolution ~\cite{Angelescu2016VLQHiggs}. The combination of these properties makes the $Zb$ channel a highly selective and powerful signal in the $\mu p$ collider environment. These properties have led the $B\to Zb$ channel to assume a central role in the search for the VLQ-$B$ quark. 

Building on the studies summarized above, it is useful to highlight several quantitative results obtained specifically for vector-like $B$ quarks at different collider facilities.
Among these, in studies on hadron colliders, single production processes in $pp$ collisions via the $B \to Zb$ channel have been extensively investigated. At this point, for the 14~TeV LHC and future higher-energy hadron machines, discovery and exclusion potentials in the $g^{\ast}$ and $M_B$ parameter spaces have been reported ~\cite{Gong2019SingleB,VLQB_pp_Zb1,Chen2017SingleB}. This literature shows that single production can become increasingly important relative to pair production, especially in the high-mass regime. It also reveals that the sensitivity to coupling parameters plays a decisive role~\cite{Atre2011SingleVLQ,Yang:2024VLQBhadronic,Han2022SingleVLQB,Han2022NuPhB975}. 

In addition, recent studies have investigated the single production of singlet vector-like $B$ quarks at future high-energy hadron colliders, such as the 100 TeV FCC-hh, particularly in the $B \to tW$ decay channel. These analyses show that masses up to the multi-TeV range can be probed for moderate coupling values, further extending the discovery potential in the high-mass regime~\cite{Han2023EPJC846}.

On the other hand, phenomenological studies on Lepton-hadron colliders have investigated the single production of singlet $B$ quarks in $ep$ machines such as LHeC and FCC--eh. These studies have reported that sensitivities up to $M_B \sim 1.4$--$2.2$~TeV can be achieved in the $B \to Zb$ and $B \to Hb$ channels~\cite{Shang2022VLQB,Gong2020SingleVLQB,ZengPanZhang2023}.

Also, in phenomenological studies on linear lepton--lepton colliders of the $e^+e^-$ type, especially in the 3~TeV CLIC scenario, single $B$ production has been investigated up to around $M_B \sim 2$~TeV~\cite{Han2022VLQCLIC,Han2024PairVLQB,Yang2025CLICPairB}.

On the other hand, on the experimental side, analyses have been performed by the ATLAS and CMS collaborations using LHC Run-2 ($\sqrt{s}=13$~TeV) data. These analyses excluded approximately the $M_B \lesssim 1.2$--$1.6$~TeV range for the singlet $B$ quark at a 95\% confidence level via the pair production channel~\cite{ATLAS:2023VLQB_single,ATLAS2024VLQpair,CMS2019VLQpair_hadronic,CMS:2022VLQpair,CMS2024VLQBsingle}. Furthermore, in single production analyses, it was shown that the obtained exclusion limits were strongly dependent on the coupling parameters. It was moreover determined that these limits were generally confined to lower mass regions~\cite{ATLAS:2023VLQB_single,CMS2024VLQBsingle}. These findings reveal that achieving a heavy-mass regime in current hadron colliders remains limited. This indicates that it is important to investigate alternative collider environments.
In this context, muon-proton ($\mu p$) colliders, especially when considered in conjunction with FCC infrastructure, stand out as a strong alternative for the search for new heavy particles thanks to their high centre of mass energies and clean initial states~\cite{Cheung:2021mup,MuonColliderReview2023, Abada2021FCCeh,Long2021MuonProton}.  High centre-of-mass energies such as $\sqrt{s}=24.5$~TeV provide an interesting environment for studying electroweak boson-initiated production mechanisms in the heavy-mass regime. In this work, this configuration is adopted as a benchmark scenario rather than as an officially established FCC design.

At this point, although various studies have investigated contact interactions, anomalous coupling structures, excited leptons, and other new-physics scenarios in the FCC--$\mu p$ environment, a combined sensitivity analysis of singlet VLQ-$B$ production through photon- and $Z$-initiated mechanisms, particularly in the $B\to Zb$ channel, has not been explored in detail.

A high-energy muon--proton collider offers a complementary environment for searching for heavy VLQs thanks to its large centre-of-mass energy and clean lepton beam. In this environment, photon- and $Z$-initiated production mechanisms provide two complementary electroweak production channels. Photon-induced production benefits from the quasi-real photon component of the muon beam, while the Z-initiated mechanism provides a complementary production mode that extends the sensitivity to different regions of the $g^\ast$--$M_B$ parameter space. The combination of these two mechanisms therefore provides a broader framework for investigating singlet VLQ-$B$ production in the multi-TeV mass region. This motivates the present sensitivity study of the $B\to Zb$ channel at the FCC--$\mu p$ collider.

Based on the significant literature mentioned above, this study investigates the single production of singlet-type VLQ-$B$ quarks through photon- and $Z$-initiated processes at the FCC--$\mu p$ collider with a centre-of-mass energy of $\sqrt{s}=24.5$~TeV. The paper is organized as follows: Section~\ref{sec:model} introduces the model framework and the effective interactions of the singlet vector-like $B$ quark. Sections~\ref{sec:photon} and~\ref{sec:zchannel} present the photon- and $Z$-initiated analyses, respectively, including the computational setup, signal process, background processes, and event selection. Section~\ref{sec:analysis} presents the common statistical framework and the corresponding sensitivity results. Finally, Section~\ref{sec:conclusion} summarizes the main results and provides concluding remarks.

\section{Model Framework}
\label{sec:model}

The theoretical framework considered in this study is based on a renormalizable, simplified extension of the SM containing an $SU(2)_L$-singlet vector-like quark, denoted by $B$, with electric charge $-1/3$. In this model, the SM field content is extended by introducing the vector-like quark whose left- and right-handed components, $B_L$ and $B_R$, transform identically under the gauge group $SU(3)_C\times SU(2)_L\times U(1)_Y$. Unlike the chiral quarks of the SM, this property allows a gauge-invariant Dirac mass term for the $B$ quark to be written independently of electroweak symmetry breaking.

The new $B$ quark can also interact with the SM quark sector through renormalizable Yukawa interactions involving the Higgs field. After electroweak symmetry breaking, the Higgs vacuum expectation value generates mixing between the vector-like $B$ quark and the SM down-type quarks. Diagonalization of the resulting mass matrix gives the physical quark mass eigenstates. As a consequence of this mixing, the heavy $B$ state acquires charged-current, neutral-current, and Higgs interactions with SM quarks.

In the simplified parametrization used here, these interactions are described by two main parameters, $g^\ast$ and $R_L$. The parameter $g^\ast$ determines the overall strength of the VLQ interactions, while $R_L$ controls the relative mixing with the first and third quark generations. With this parametrization, the interactions of the singlet vector-like $B$ quark with SM fermions and electroweak bosons are described by the following Lagrangian~\cite{VLQ_review2,Buchkremer2013VLQUFO,VLQB_pp_Zb1}: \begin{align}
\mathcal{L}_{\text{int}}
&=
\Bigg(\frac{g\,g^\ast}{\sqrt{2}}\Bigg)
\Bigg[
\sqrt{\frac{R_L}{1+R_L}}\frac{1}{\sqrt{2}}\,\bar{B}_L\gamma^\mu W^-_\mu u_L
\nonumber \\
&\quad
+
\sqrt{\frac{1}{1+R_L}}\frac{1}{\sqrt{2}}\,\bar{B}_L\gamma^\mu W^-_\mu t_L
\nonumber \\
&\quad
+
\sqrt{\frac{R_L}{1+R_L}}\frac{1}{2\cos\theta_W}\,\bar{B}_L\gamma^\mu Z_\mu d_L
\nonumber \\
&\quad
+
\sqrt{\frac{1}{1+R_L}}\frac{1}{2\cos\theta_W}\,\bar{B}_L\gamma^\mu Z_\mu b_L
\nonumber \\
&\quad
-
\sqrt{\frac{R_L}{1+R_L}}\frac{M_B}{2m_W}\,\bar{B}_R H d_L
\nonumber \\
&\quad
-
\sqrt{\frac{1}{1+R_L}}\frac{M_B}{2m_W}\,\bar{B}_R H b_L
\Bigg]
+\text{h.c.}
\end{align}

where $g$ is the $SU(2)_L$ weak coupling constant, $g^\ast$ is the effective coupling parameter associated with the new physics, $\theta_W$ is the Weinberg angle, $M_B$ is the mass of the vector-like $B$ quark, and $m_W$ is the mass of the $W$ boson. The fields $u_L$, $d_L$, $t_L$, and $b_L$ denote the left-handed components of the SM up, down, top, and bottom quarks, respectively; $W^-_\mu$ and $Z_\mu$ denote the electroweak gauge bosons; and $H$ represents the physical Higgs boson field. Here, ``h.c.'' denotes the Hermitian conjugate terms.

In this parametrisation, $R_L$ controls the relative strength of the interactions involving the first and third generations. In the adopted UFO implementation, the corresponding left-handed flavour weights are defined as
\begin{equation}
\zeta^L_{Bd}=\frac{R_L}{1+R_L}, \qquad
\zeta^L_{Bb}=\frac{1}{1+R_L}, \qquad
\zeta^L_{Bs}=0.
\end{equation}

It should be stressed that $R_L$ and the SM CKM matrix are treated as independent flavour inputs in the \texttt{BsingletVL\_UFO} implementation. While $R_L$ determines the VLQ flavour weights given above, the CKM matrix enters independently through the standard SM charged-current vertices.

Thus, the first- and third-generation VLQ interactions are controlled by $\zeta_{Bd}^{L}$ and $\zeta_{Bb}^{L}$, respectively, while direct VLQ mixing with the second generation is set to zero. For the charged-current $WBq$ interaction, the dominant light-quark-initiated contribution arises from the first-generation up-quark channel. Accordingly, the photon- and $Z$-initiated production mechanisms are dominated by $\gamma u\to W^+B$ and $Zu\to W^+B$, respectively. A charm-quark-initiated contribution can also be present in the full $Z$-initiated process. However, its contribution is numerically very small compared with the dominant up-quark channel and does not affect the sensitivity results. 
The limiting behaviour of this parametrisation is straightforward. For $R_L\to0$, the first-generation mixing weight $\zeta_{Bd}^{L}$ vanishes, while $\zeta_{Bb}^{L}\to1$, so that the VLQ-$B$ interactions become third-generation dominated and the dominant light-quark-initiated single-production contributions considered in this work are strongly suppressed. Conversely, increasing $R_L$ enhances the relative contribution of the first-generation interaction. Separately, for $g^\ast\to0$, the mixing-induced interactions of the VLQ-$B$ with SM quarks through the $W$, $Z$, and Higgs bosons vanish. The electromagnetic and strong gauge interactions of the VLQ itself are unaffected by this limit.

For a fixed value of $R_L$, the branching fractions of the singlet VLQ-$B$ are determined by the effective Lagrangian and are largely independent of $g^\ast$, since the corresponding partial decay widths scale as $(g^\ast)^2$. Consequently, $g^\ast$ controls the overall strength of the mixing-induced interactions and the total decay width, whereas $R_L$ determines the relative contributions of the first- and third-generation channels. The total decay width $\Gamma_B$ increases with both $M_B$ and $g^\ast$; therefore, the relative width $\Gamma_B/M_B$ is explicitly considered when presenting the explored parameter space.

For the small but non-zero benchmark values $R_L=0.05$ and $R_L=0.01$ considered in this work, the interactions remain predominantly third-generation dominated while they retain suppressed first-generation contribution. In the heavy-mass limit, the dominant third-generation decay modes approach the characteristic singlet pattern
$\mathrm{BR}(B\to Wt):\mathrm{BR}(B\to Zb):\mathrm{BR}(B\to Hb)\simeq2:1:1$,
in accordance with the Goldstone-boson equivalence behaviour. Accordingly, the branching fractions are not treated as independent free parameters but are fixed by the chosen $(g^\ast,R_L,M_B)$ benchmark scenario.

Compared with doublet or higher-dimensional VLQ representations, which generally introduce additional heavy fermionic states and interaction structures, the singlet scenario provides a minimal framework for studying the production and decay phenomenology of a single vector-like $B$ quark. The singlet representation is therefore adopted here as a clean benchmark for investigating the sensitivity to photon- and $Z$-initiated single VLQ-$B$ production.

\begin{figure*}[t]
    \centering
    \includegraphics[width=0.78\textwidth]{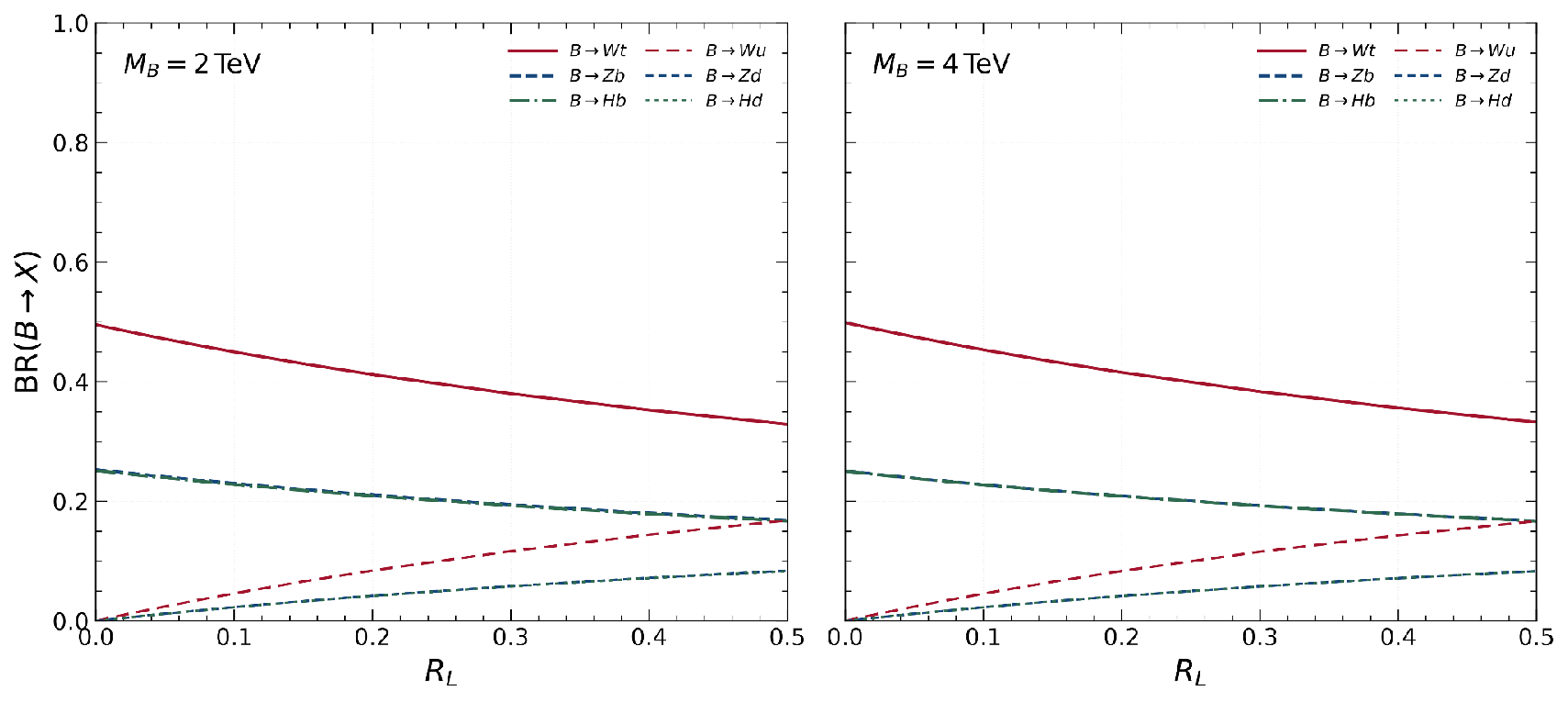}
    \caption{Branching ratios of the singlet vector-like $B$ quark as a function of $R_L$ for $M_B=2$ and $4~\mathrm{TeV}$ at $g^\ast=0.2$.}
    \label{fig:br_rl}
\end{figure*}

\subsection{Effective Photon and $Z$-Boson Descriptions}

At multi-TeV energies, energetic leptons can radiate electroweak gauge bosons predominantly in the forward and collinear regions. This feature allows a high-energy muon beam to act effectively as a source of neutral electroweak gauge bosons and provides an important class of production mechanisms for heavy particles in muon-based collider environments~\cite{Ruiz2022EVA}. In the present study, the photon-induced and $Z$-initiated channels are described using different effective approaches appropriate to the corresponding gauge bosons. The quasi-real photon spectrum emitted from the incoming muon is treated within the improved Weizs\"acker--Williams (IWW) approximation~\cite{Frixione1993IWW}, whereas the effective $Z$-boson content of the muon beam is described within the effective vector-boson approximation (EVA)~\cite{Ruiz2022EVA}.

In the photon-induced channel, the incoming muon radiates a predominantly quasi-real photon before the hard interaction with a parton from the proton. Because the photon is massless, its emission is strongly enhanced in the small-angle and collinear region, providing the physical basis of the IWW description~\cite{Frixione1993IWW}. The emitted photon then participates in the hard scattering, while the parent muon is expected to retain a large fraction of its initial longitudinal momentum and continue predominantly in the forward direction.

For the $Z$-initiated channel, the incoming muon radiates a neutral weak gauge boson before the hard interaction. In contrast to the quasi-real photon, the $Z$ boson is massive, and its effective distribution is therefore modified by the electroweak mass scale. Nevertheless, at multi-TeV muon energies, effective $Z$ emission can provide a relevant contribution to hard scattering processes~\cite{Ruiz2022EVA}. In this regime, EVA provides a useful framework for describing the approximately collinear emission of massive electroweak gauge bosons from the incoming muon.

It is evident that a common feature of both mechanisms is the presence of a surviving muon associated with the gauge-boson-emission process. It is expected that this muon will remain predominantly in the forward region and will carry a significant proportion of the beam momentum. High-energy muon collider studies have indicated that forward-muon tagging may offer an additional method for identifying neutral vector-boson-initiated processes ~\cite {Ruhdorfer2025ForwardMuon}. Although forward-muon tagging is not explicitly included in the event selection of the present parton-level analysis, it may provide an additional experimental signature in a realistic detector environment. It is therefore evident that the two mechanisms differ in their effective gauge-boson spectra, production rates, and kinematic properties. However, both provide access to single production of the singlet VLQ-$B$.
It should be noted that the effective-boson descriptions adopted here treat the photon-induced and $Z$-initiated contributions separately by construction and therefore do not retain amplitude-level $\gamma$--$Z$ interference~\cite{Ruiz2022EVA}. As an additional check, we examined the interference at the hard-matrix-element level for the transverse $\gamma$- and $Z$-initiated subprocesses. The interference term, $2\,\mathrm{Re}(\mathcal{M}_{\gamma}\mathcal{M}_{Z}^{*})$, was found to be non-zero and constructive for the representative kinematic configurations examined. Since this check does not correspond to a fully coherent $\mu p$-level convolution of the IWW photon and EVA $Z$ contributions, the photon- and $Z$-initiated channels are treated and presented separately in the phenomenological analysis.

In this regard, both channels are investigated through the same $B\to Zb$ decay mode and lead to the same reconstructed $\ell^{+}\ell^{-}bjj$ topology from the hard process. Their different electroweak radiation patterns and production dynamics therefore provide complementary sensitivity to the single production of a heavy singlet VLQ-$B$ at a high-energy muon--proton collider.

\section{Photon-Induced VLQ-$B$ Analysis}
\label{sec:photon}

\subsection{Computational Setup}

The photon-induced analysis is performed for the FCC--$\mu p$ collider configuration in which a $3$~TeV muon beam collides with a $50$~TeV proton beam, corresponding to a centre-of-mass energy of $\sqrt{s}=24.5$~TeV. The production cross sections for the signal and SM background processes were calculated at parton level using \textsc{MadGraph5\_aMC@NLO} v3.6.3~\cite{MG5}. The interactions of the vector-like $B$ quark were implemented through the \texttt{BsingletVL\_UFO} model~\cite{Degrande2012UFO,Alloul2014FeynRules,VLQB_pp_Zb1,Buchkremer2013VLQUFO,VLQBsingletUFO}. The proton structure was described using the \texttt{NNPDF3.1 NNLO} parton distribution function set with LHAPDF ID 303600~\cite{NNPDF31}.

For the photon-induced channel, the signal process was generated using the following command
\begin{lstlisting}[breaklines=true,basicstyle=\ttfamily\small]
generate a p > w+ bp, (bp > z b, z > l+ l-), (w+ > j j)
\end{lstlisting}
The photon spectrum emitted from the incoming muon was described using the improved Weizs\"acker--Williams (IWW) approximation as implemented in \textsc{MadGraph5}\allowbreak\textsc{\_aMC}\allowbreak\textsc{@NLO}~\cite{Frixione1993IWW}. The beam configuration was specified by \texttt{lpp1=4}, \texttt{lpp2=1}, \texttt{pdlabel1=iww}, and \texttt{pdlabel2}\allowbreak\texttt{=lhapdf}. The maximum photon virtuality was set to
$Q_{\max}^{2}=2~\mathrm{GeV}^{2}$. 

The signal parameter space was scanned over $1500~\mathrm{GeV}\leq M_B\leq3000~\mathrm{GeV}$ and $0.1\leq g^\ast\leq0.8$, with mass and coupling steps of $250$~GeV and $0.05$, respectively. The direct signal grid was generated for the benchmark mixing value $R_L=0.05$ and $R_L=0.01$. The total width of the vector-like $B$ quark was evaluated automatically for each benchmark point. No analysis-level selection cuts were imposed at the event-generation stage.

\subsection{Signal Process and Final State}

\begin{figure}[t]
    \centering
    \includegraphics[width=\linewidth]{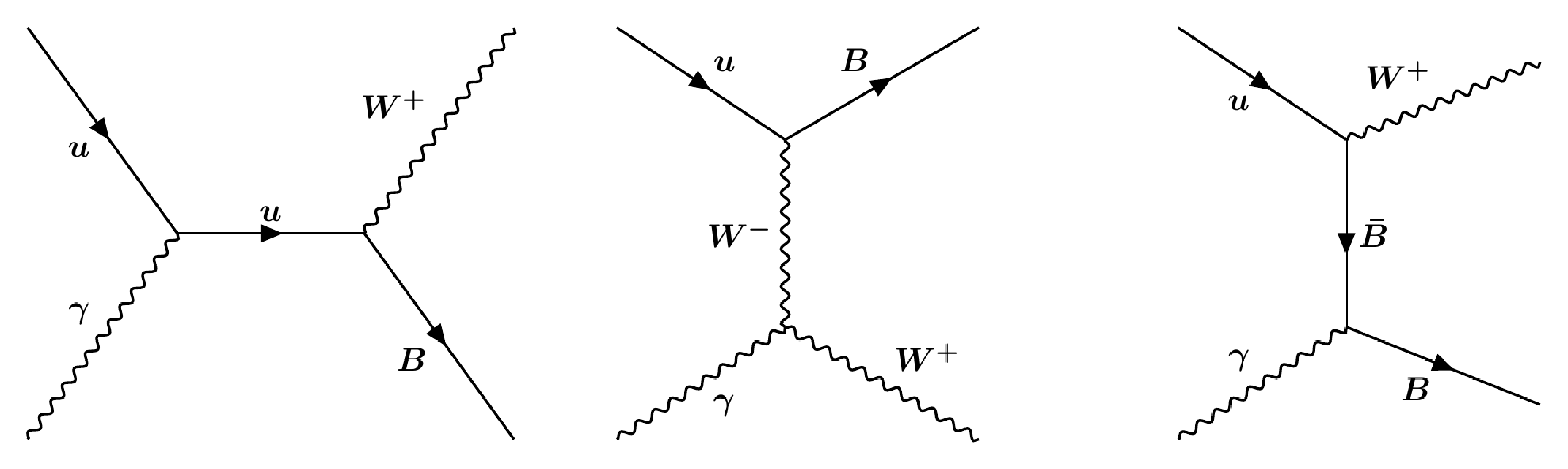}
    \caption{Representative tree-level Feynman diagrams for the photon-induced subprocess $\gamma u\to W^{+}B$.}
    \label{fig:feynman_subprocess}
\end{figure}

In the photon-induced channel, the single production of the vector-like $B$ quark proceeds through the interaction of an equivalent photon radiated from the incoming muon with an up quark from the proton. The corresponding hard subprocess is
\begin{equation}
\gamma u \rightarrow W^{+}B.
\end{equation}
Representative tree-level Feynman diagrams for this production mechanism are shown in Fig.~\ref{fig:feynman_subprocess}.

At the muon--proton collision level, the signal process can be written as
\begin{equation}
\mu p \rightarrow \mu W^{+}BX \rightarrow \mu W^{+}(Zb)X \rightarrow \mu \ell^{+}\ell^{-}bjjX,
\end{equation}
where $\ell=e,\mu$ and $X$ denotes the proton remnant. In this analysis, the vector-like $B$ quark is reconstructed through the decay $B\to Zb$, with the $Z$ boson decaying leptonically via $Z\to\ell^{+}\ell^{-}$ and the accompanying $W$ boson decaying hadronically through $W\to jj$. The reconstructed hard final state therefore consists of two oppositely charged leptons, one $b$ jet, and two light jets, while the surviving muon remains associated with the photon-emission leg and is expected predominantly in the forward region.

The leptonic decay of the $Z$ boson provides a clean dilepton signature and allows an accurate reconstruction of the $Z$-boson invariant mass, while the hadronic decay of the $W$ boson provides additional information on the production topology. These features make the $\ell^{+}\ell^{-}bjj$ final state particularly suitable for probing a heavy vector-like $B$ quark in the photon-induced channel.

Accordingly, the total cross section for photon-induced single-$B$ production in muon--proton collisions can be expressed as
\begin{multline}
\sigma(\mu p \to \mu W^{+}BX)
=
\int dx_\gamma \int dx_u\,
f_{\gamma/\mu}(x_\gamma) \\
\times f_{u/p}(x_u,\mu_F)\,
\hat{\sigma}(\gamma u \to W^{+}B),
\label{eq:sigma_mup_WB}
\end{multline}
where $f_{\gamma/\mu}(x_\gamma)$ denotes the effective photon distribution of the incoming muon, $f_{u/p}(x_u,\mu_F)$ is the up-quark distribution in the proton, and $\hat{\sigma}(\gamma u\to W^{+}B)$ represents the corresponding partonic cross section.

\subsection{Production Cross Section}

\begin{figure}[t]
    \centering
    \includegraphics[width=0.48\textwidth]{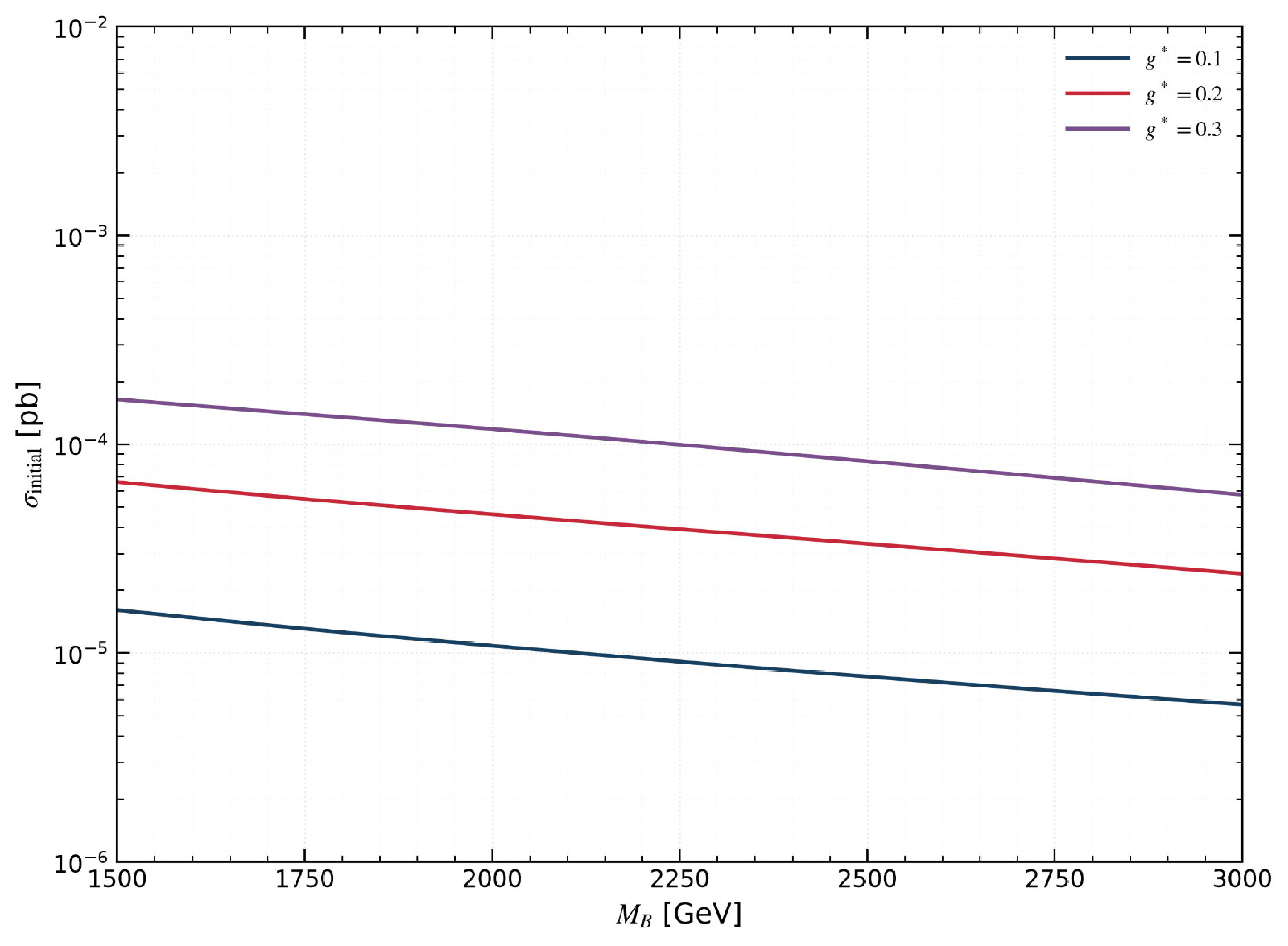}
     \caption{Initial cross section of the photon-induced signal process, including the decay chain $B\to Zb$, $Z\to \ell^+\ell^-$, and $W\to jj$, as a function of the vector-like $B$-quark mass for different   values of the coupling parameter $g^\ast$.}
    \label{fig:sigma_vs_MB}
\end{figure}

In this context, the initial cross section of the photon-induced signal process, including the decay chain $B\to Zb$, $Z\to \ell^+\ell^-$, and $W\to jj$, was calculated as a function of the vector-like $B$-quark mass $M_B$ for different values of the coupling parameter $g^\ast$, and the results are presented in Fig.~\ref{fig:sigma_vs_MB}. As can be seen from the figure, the signal cross section decreases monotonically as $M_B$ increases. Furthermore, larger values of $g^\ast$ significantly enhance the signal rate over the scanned mass range.

\begin{figure}[t]
    \centering
    \includegraphics[width=0.52\textwidth]{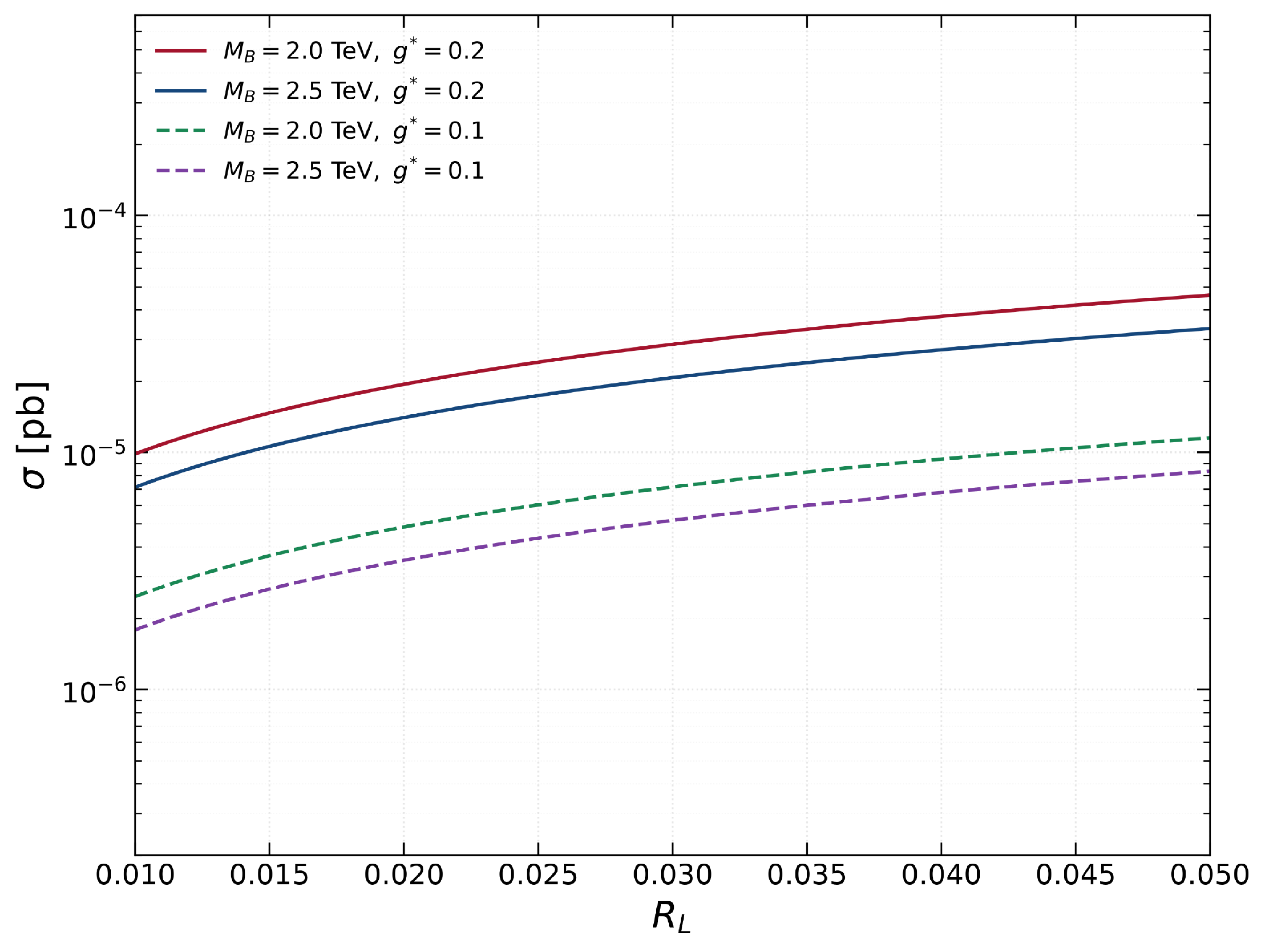}
    \caption{Photon-induced signal production cross section as a function of
    $R_L$ for $M_B=2.0$ and $2.5~\mathrm{TeV}$, shown for the benchmark
    coupling values $g^\ast=0.1$ and $0.2$.}
    \label{fig:photon_sigma_RL}
\end{figure}

In addition to the $M_B$ and $g^\ast$ dependence, the signal rate is also sensitive to the first-generation mixing parameter $R_L$. This dependence is illustrated in Fig.~\ref{fig:photon_sigma_RL}, where the photon-induced signal cross section is shown as a function of $R_L$ for representative values of $M_B$ and $g^\ast$. The cross section increases with $R_L$, reflecting the enhanced first-generation contribution to the production of the vector-like $B$ quark.

For a fixed value of $M_B$, the photon-induced signal cross section approximately follows the expected $\sigma\propto(g^\ast)^2$ behaviour in the narrow-width regime. As $g^\ast$ increases, however, the total width $\Gamma_B$ also grows, and finite-width effects become increasingly relevant. As a result, deviations from a simple quadratic coupling dependence can appear in the broader-width region. The resonance-width behaviour is therefore taken into account when interpreting the corresponding sensitivity regions.

\subsection{SM Backgrounds}
\label{sec:backgrounds}

To identify the dominant SM processes that can mimic the photon-induced signal, background channels with the same or closely related reconstructed final-state topology are considered. In the present analysis, the dominant photon-induced SM backgrounds are taken to be
\begin{equation}
\gamma p \to WZj, \qquad
\gamma p \to WZb, \qquad
\gamma p \to ZZj.
\end{equation}

The corresponding samples were generated at parton level using the following commands
\begin{lstlisting}[
  breaklines=true,
  breakatwhitespace=true,
  basicstyle=\ttfamily\footnotesize
]
generate a p > w+ z j, (w+ > j j), (z > l+ l-)
generate a p > w+ z b, (w+ > j j), (z > l+ l-)
generate a p > z z j, (z > l+ l-), (z > j j)
\end{lstlisting}
for the $WZj$, $WZb$, and $ZZj$ processes, respectively. In addition to these dominant backgrounds, we also explicitly examined the photon-induced $Zjjj$ and $Zb\bar{b}j$ processes, which can in principle contribute to the selected final state through light-jet mistagging or additional heavy-flavour production. These samples were analysed using the same IWW configuration, detector-smearing prescription, tagging treatment, and event-selection criteria as those applied to the other photon-induced backgrounds. Although both processes retain non-zero rates after the intermediate reconstruction requirements, no generated events survive the final $M_{b\ell\ell}$ selection. Their contributions are therefore negligible for the photon-channel sensitivity analysis. Other photon-induced processes may also contribute after particle decays and event reconstruction. The $t\bar{t}q$-type contribution is especially suppressed by the requirement of a same-flavour, opposite-sign dilepton pair compatible with an on-shell $Z$ boson, together with the high-$M_{b\ell\ell}$ requirement. The additional neutrinos from semileptonic top decays further reduce its overlap with the selected signal topology. Moreover, contributions from non-photon-induced production mechanisms are not included in the present analysis, which focuses specifically on the photon-induced channel. In an experimental setting, the forward-scattered muon associated with quasi-real photon emission may in principle provide an additional handle for distinguishing photon-induced events from other production mechanisms.

The incoming photon is described using the same IWW setup adopted for the signal process. These background samples are analysed using the same detector-smearing and object-reconstruction procedure as the photon-induced signal. The same kinematic selection strategy is then applied in order to suppress the SM background contributions while retaining the characteristic signal topology.

\subsection{Event Selection and Detector Smearing}

\begin{figure*}[t]
\centering
\begin{minipage}{0.48\textwidth}
    \centering
    \includegraphics[width=\textwidth]{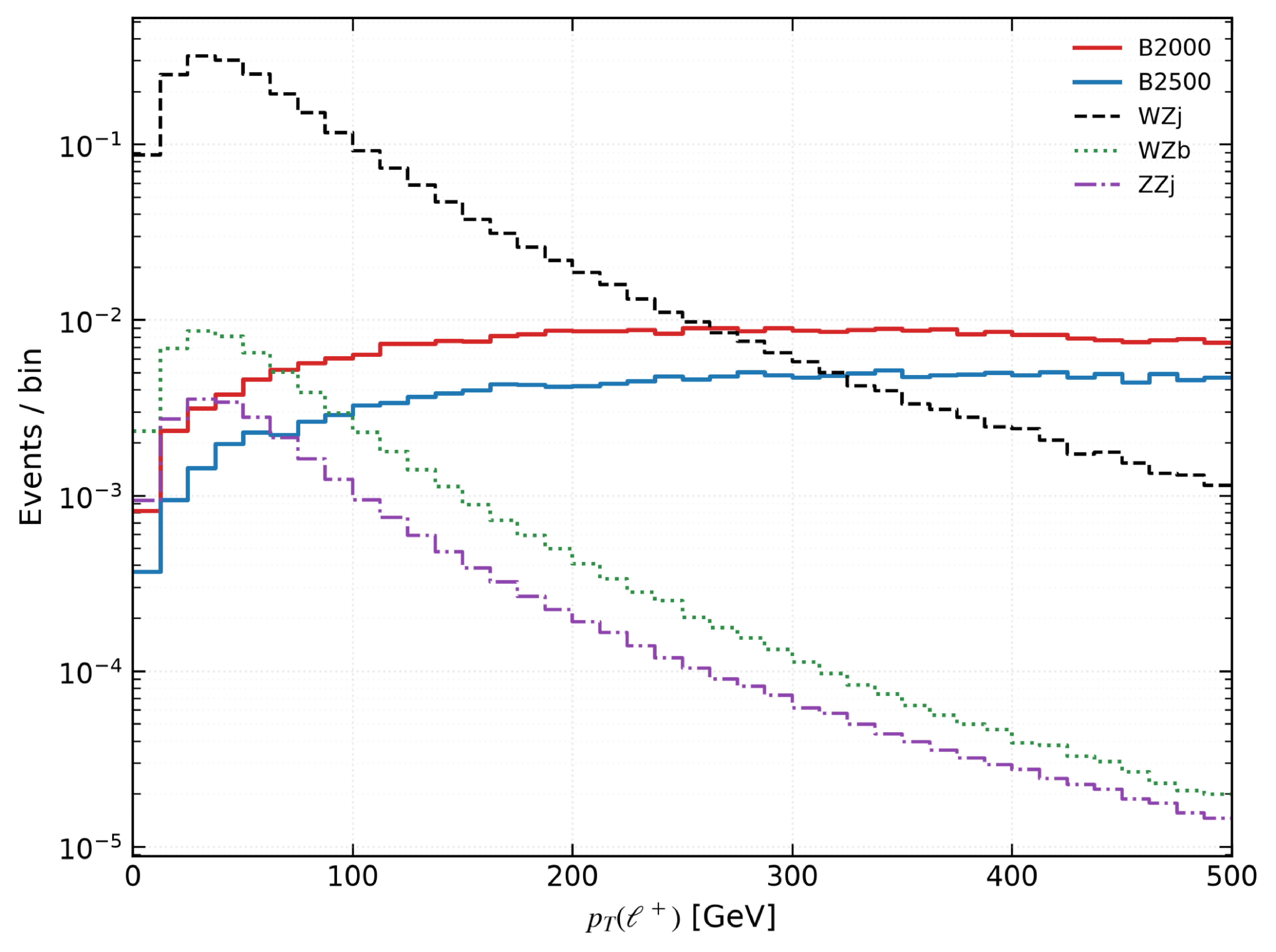}
    \par\small (a) $p_T(\ell)$ distribution
\end{minipage}
\hfill
\begin{minipage}{0.48\textwidth}
    \centering
    \includegraphics[width=\textwidth]{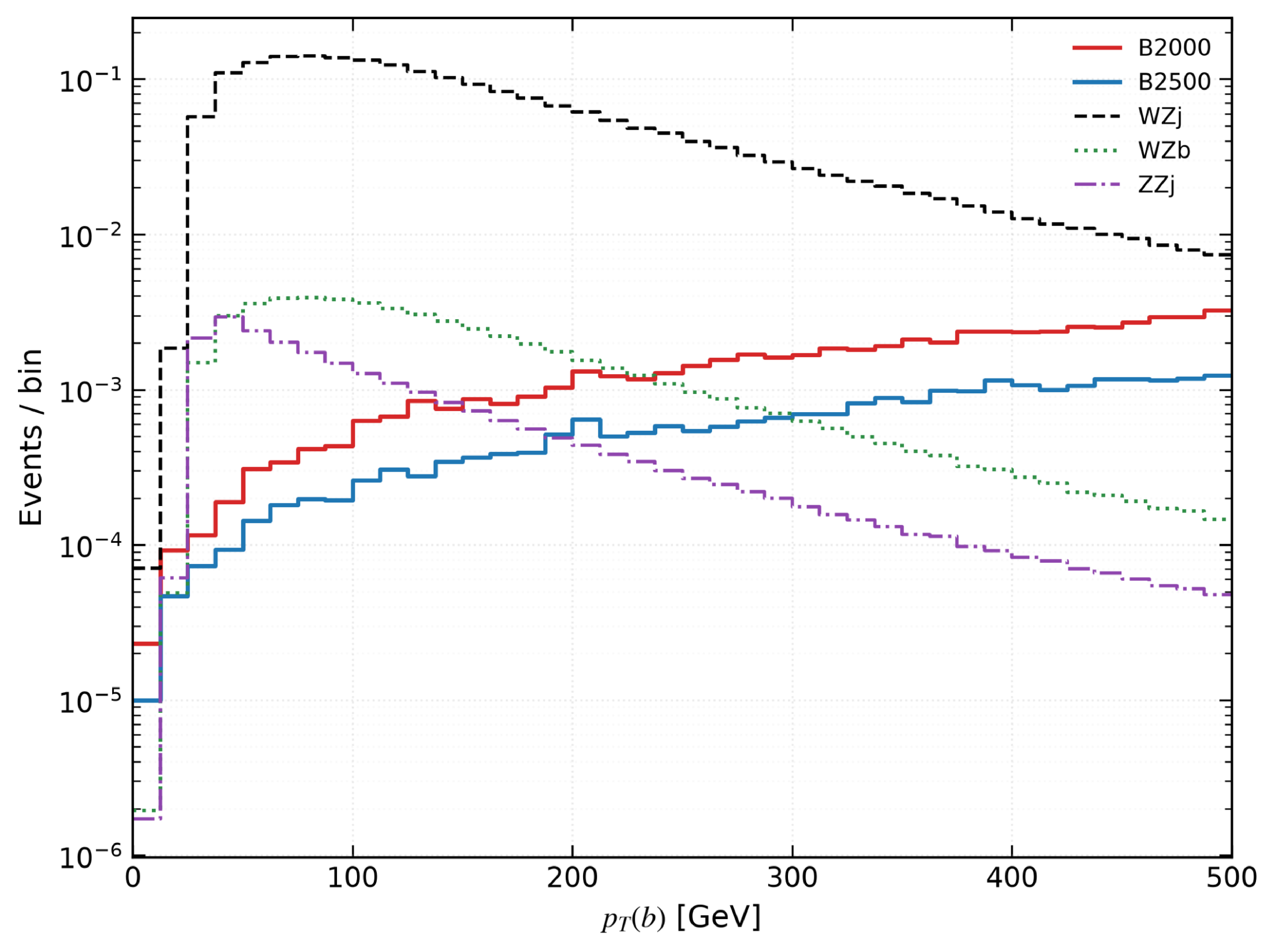}
    \par\small (b) $p_T(b)$ distribution
\end{minipage}

\vspace{0.4cm}

\begin{minipage}{0.48\textwidth}
    \centering
    \includegraphics[width=\textwidth]{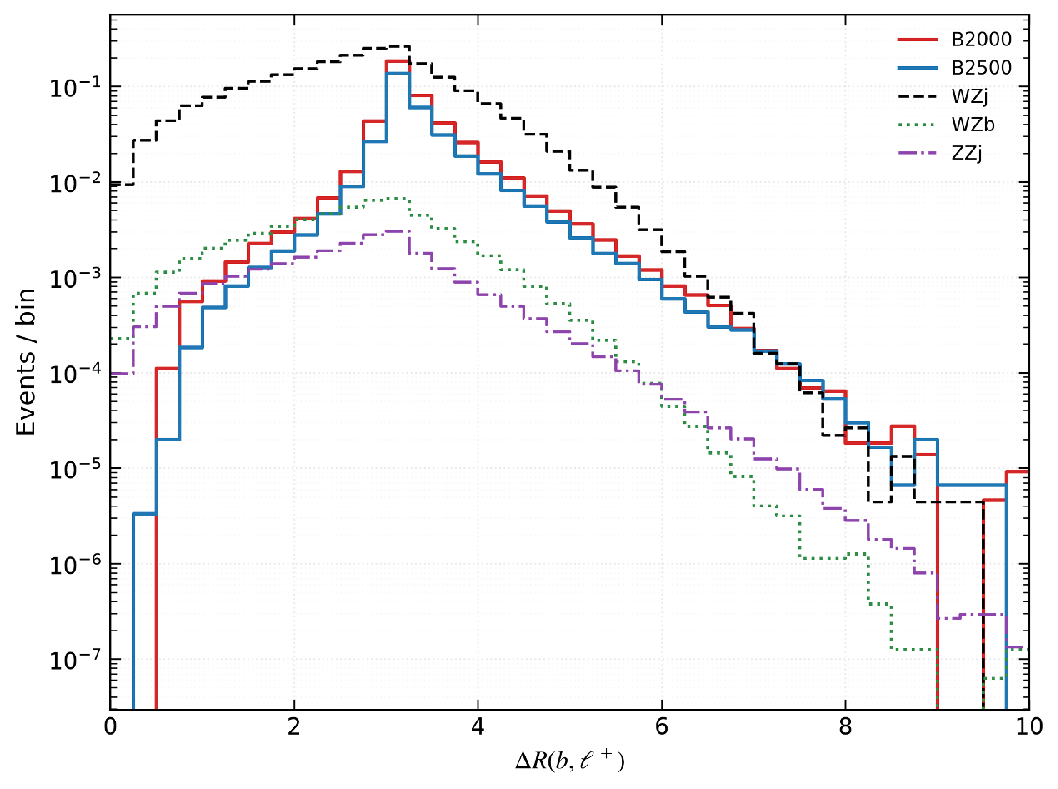}
    \par\small (c) $\Delta R(b,\ell)$ distribution
\end{minipage}
\hfill
\begin{minipage}{0.48\textwidth}
    \centering
    \includegraphics[width=\textwidth]{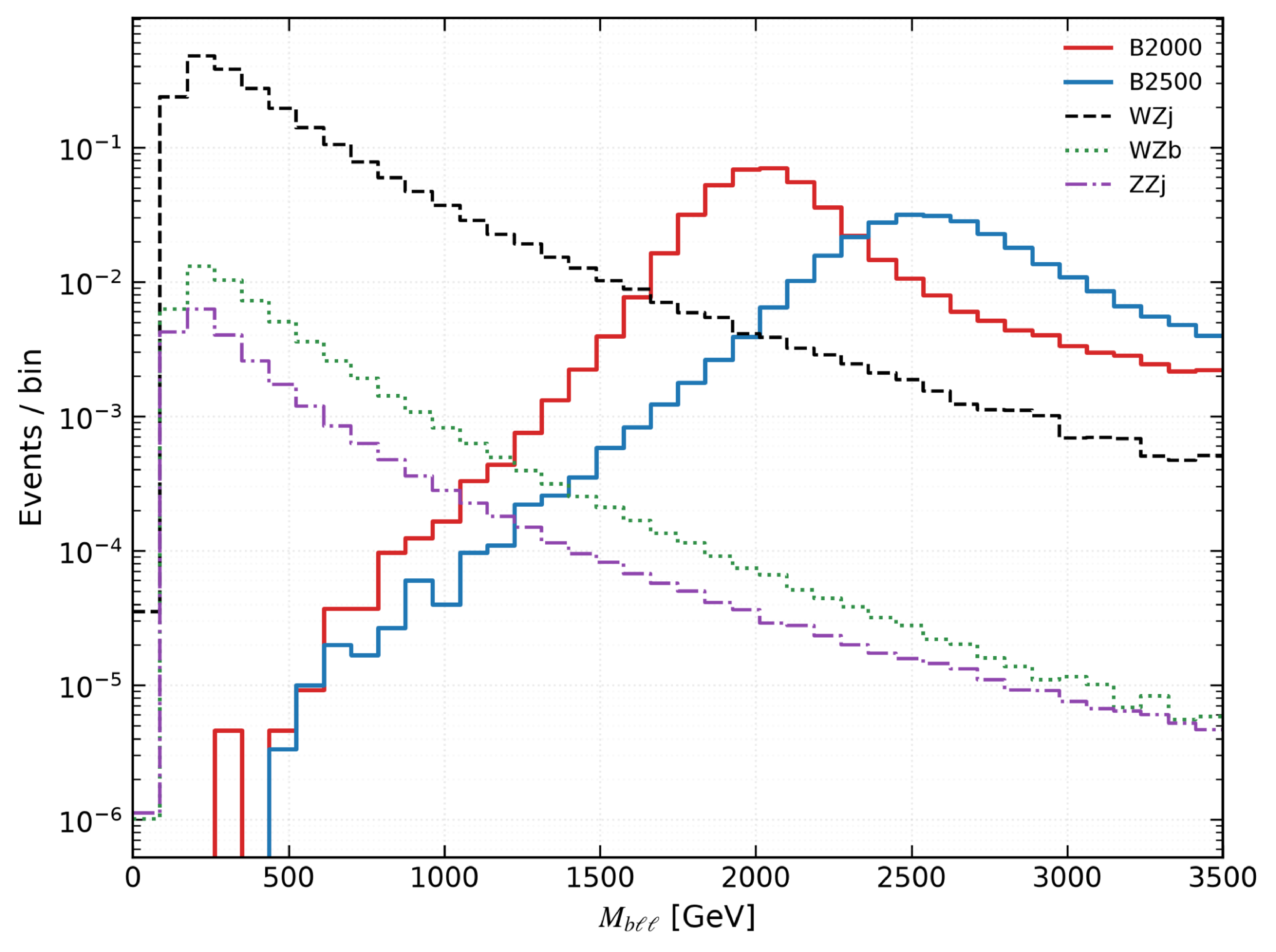}
    \par\small (d) $M_{b\ell\ell}$ distribution
\end{minipage}

\caption{Initial smeared kinematic distributions of the variables used in the analysis for photon induced  signal and background processes. The signal distributions are shown for the benchmark coupling value $g^\ast = 0.2$. The event yields are normalised to an integrated luminosity of $\mathcal{L}=10~\mathrm{fb^{-1}}$.}
\label{fig:kinematic_overview}
\end{figure*}

Accordingly, the distributions of several representative kinematic variables are presented in Fig.~\ref{fig:kinematic_overview}. As shown in Fig.~\ref{fig:kinematic_overview}(a) and (b), the leptons and $b$ jets in the signal tend to populate higher transverse-momentum ($p_T$) regions than the corresponding background processes. This behaviour reflects the production and decay of the heavy $B$ resonance, which leads to more energetic final-state objects. In contrast, the low-$p_T$ region is dominated mainly by the $WZj$ and $ZZj$ backgrounds. This separation motivates the use of a lower transverse-momentum requirement of $p_T>20~\mathrm{GeV}$ as an initial object-selection criterion.

Figure~\ref{fig:kinematic_overview}(c) shows the distribution of the angular separation $\Delta R(b,\ell)$. The signal events are concentrated mainly in the region $3\lesssim\Delta R(b,\ell)\lesssim4$, whereas the background processes exhibit broader distributions. This difference indicates that $\Delta R(b,\ell)$ provides useful discrimination between the signal and SM background processes.

The invariant-mass distribution $M_{b\ell\ell}$ is shown in Fig.~\ref{fig:kinematic_overview}(d). For the signal, a clear resonance structure appears around the vector-like $B$-quark mass, whereas the SM backgrounds exhibit smooth and continuously decreasing distributions. The $M_{b\ell\ell}$ variable therefore provides a direct reconstruction of the heavy $B$ resonance and constitutes one of the most effective observables for separating the signal from the background.

Based on these kinematic features, the event-selection requirements were applied sequentially to enhance the signal-to-background separation, as described below.

\begin{description}

\item[\textbf{Cut--1:}]
For detector acceptance and basic object selection,
\begin{itemize}
\setlength{\itemsep}{4pt}
\item $p_T^\ell > 20~\mathrm{GeV}$, $|\eta^\ell| < 2.5$,
\item $p_T^b > 20~\mathrm{GeV}$, $|\eta^b| < 4.0$,
\item $p_T^j > 20~\mathrm{GeV}$, $|\eta^j| < 5.0$.
\end{itemize}

\item[\textbf{Cut--2 (Angular Separation):}]
\[
3.2 < \Delta R(b,\ell) < 3.8
\]

\item[\textbf{Cut--3 (Boson Reconstruction):}]
\[
75~\mathrm{GeV} < M_{jj} < 85~\mathrm{GeV}, \qquad
85~\mathrm{GeV} < M_{\ell\ell} < 95~\mathrm{GeV}
\]

\item[\textbf{Cut--4 (Heavy $B$ Reconstruction):}]
\[
M_{\ell\ell b} > 1500~\mathrm{GeV}
\]

\end{description}

To assess the impact of detector-resolution effects on the analysis, a simplified detector-resolution study was performed by applying Gaussian smearing to the parton-level final-state objects before the analysis selections. Such Gaussian smearing procedures are commonly used in phenomenological studies to approximate detector-resolution effects~\cite{Araz:2021,Buckley:2019}. In particular, jet-energy resolution can noticeably affect reconstructed observables such as invariant-mass distributions~\cite{Cassidy:2024}.

In this approach, the kinematic variables of the final-state objects are modified by introducing relative Gaussian fluctuations. The transverse momentum of charged leptons is smeared according to
\begin{equation}
p_T^{\mathrm{smeared}}
=
p_T(1+\delta_\ell),
\qquad
\delta_\ell\sim\mathcal{N}(0,\sigma_\ell),
\end{equation}
while the energies of light jets are smeared as
\begin{equation}
E^{\mathrm{smeared}}
=
E(1+\delta_j),
\qquad
\delta_j\sim\mathcal{N}(0,\sigma_j),
\end{equation}
and for $b$ jets,
\begin{equation}
E^{\mathrm{smeared}}
=
E(1+\delta_b),
\qquad
\delta_b\sim\mathcal{N}(0,\sigma_b).
\end{equation}

The resolution parameters are chosen as $\sigma_\ell=0.01$ for charged leptons, $\sigma_j=0.10$ for light jets, and $\sigma_b=0.12$ for $b$ jets, reflecting their different detector responses. The angular variables $\eta$ and $\phi$ are kept unchanged. For background processes without a true $b$ quark, such as $WZj$ and $ZZj$, a light-jet-to-$b$ mistag probability of $1\%$ is included.

The quantitative effects of the sequential event-selection requirements on the signal and background processes are summarized in Table~\ref{tab:cutflow_smear}. At the initial stage, the background cross sections are several orders of magnitude larger than the signal cross section. After the angular-separation, boson-reconstruction, and high-invariant-mass requirements are applied, the background contributions are strongly reduced. At the same time, approximately $10$--$11\%$ of the signal events are retained for the representative benchmark masses considered. These results show that the adopted selection strategy provides an effective separation between the VLQ-$B$ signal and the dominant SM backgrounds.

\begin{table*}[htbp]
\centering
\caption{Cut-flow cross sections for the photon induced signal and background processes after
successive selection criteria. The signal results are shown for the benchmark
coupling $g^*=0.2$ and two representative VLQ masses. Detector-level energy
smearing is applied throughout the cut-flow. All cross sections are given in pb.}
\label{tab:cutflow_smear}
\begin{tabular}{lccccc}
\hline
Selection
& Signal ($M_B=2000~\mathrm{GeV}$)
& Signal ($M_B=2500~\mathrm{GeV}$)
& $WZj$
& $WZb$
& $ZZj$ \\
\hline

Initial
& $4.614\times10^{-5}$
& $3.333\times10^{-5}$
& $2.204\times10^{-4}$
& $5.649\times10^{-6}$
& $2.400\times10^{-6}$ \\

Cut 1
& $2.764\times10^{-5}$
& $2.186\times10^{-5}$
& $1.094\times10^{-4}$
& $2.811\times10^{-6}$
& $1.097\times10^{-6}$ \\

Cut 2
& $9.500\times10^{-6}$
& $7.715\times10^{-6}$
& $1.789\times10^{-5}$
& $4.556\times10^{-7}$
& $1.551\times10^{-7}$ \\

Cut 3
& $4.516\times10^{-6}$
& $3.731\times10^{-6}$
& $8.516\times10^{-6}$
& $2.175\times10^{-7}$
& $2.250\times10^{-8}$ \\

Cut 4
& $4.475\times10^{-6}$
& $3.720\times10^{-6}$
& $6.722\times10^{-7}$
& $1.029\times10^{-8}$
& $1.139\times10^{-9}$ \\

Efficiency (\%)
& $9.70$
& $11.16$
& $3.05\times10^{-1}$
& $1.82\times10^{-1}$
& $4.74\times10^{-2}$ \\

\hline
\end{tabular}
\end{table*}

\section{$Z$-Initiated VLQ-$B$ Analysis}
\label{sec:zchannel}

\subsection{Computational Setup}

The $Z$-initiated analysis is performed using the same FCC--$\mu p$ collider configuration, \textsc{MadGraph5}\allowbreak\textsc{\_aMC}\allowbreak\textsc{@NLO} version, \texttt{BsingletVL\_UFO} model, and NNPDF3.1 NNLO proton parton distribution functions as in the photon-induced analysis.

For the signal generation, the $Z$-initiated process was generated using the following command
\begin{lstlisting}[
  breaklines=true,
  breakatwhitespace=true,
  basicstyle=\ttfamily\footnotesize
]
generate z p > w+ bp, (bp > z b, z > l+ l-), (w+ > j j)
\end{lstlisting}
The effective $Z$-boson content of the incoming muon beam was described within the effective vector-boson approximation (EVA) as implemented in \textsc{MadGraph5}\allowbreak\textsc{\_aMC}\allowbreak\textsc{@NLO}. The corresponding beam setup was defined by \texttt{lpp1=4}, \texttt{lpp2=1}, \texttt{pdlabel1=eva},\allowbreak \texttt{pdlabel2=lhapdf}, and \texttt{ievo\_eva=0}. The factorization and renormalization scales were evaluated dynamically using \texttt{fixed\_fac\_scale1}\allowbreak\texttt{=False}, \texttt{fixed\_fac\_scale2=False}, \texttt{fixed\_ren\_scale=False}, \texttt{dynamical\_scale\_choice=4}, and \texttt{scalefact=1.0}.

For the $Z$-initiated signal analysis, the parameter space was scanned over $M_B=1500$--$4500~\mathrm{GeV}$ in steps of $250~\mathrm{GeV}$ and $0.02\leq g^\ast\leq0.50$. A finer coupling scan was performed in the low-$g^\ast$ region, where the sensitivity contours are most relevant, while larger coupling values were sampled more coarsely. Independent signal grids were generated for the benchmark mixing values $R_L=0.05$ and $R_L=0.01$ in order to investigate the dependence of the sensitivity on the mixing parameter. The generated signal samples were subjected to the same detector-smearing prescription, followed by the event-selection requirements defined for the $Z$-initiated analysis.

The total width of the vector-like $B$ quark was evaluated for each generated parameter point. The resulting ratios $\Gamma_B/M_B$ were used to construct the $1\%$, $10\%$, and $20\%$ width contours shown together with the discovery and exclusion sensitivities. These contours indicate the resonance-width regime associated with the explored regions of the $g^\ast$--$M_B$ parameter space.

\subsection{Signal Process and Final State}

\begin{figure*}[t]
    \centering
    \includegraphics[width=0.95\textwidth]{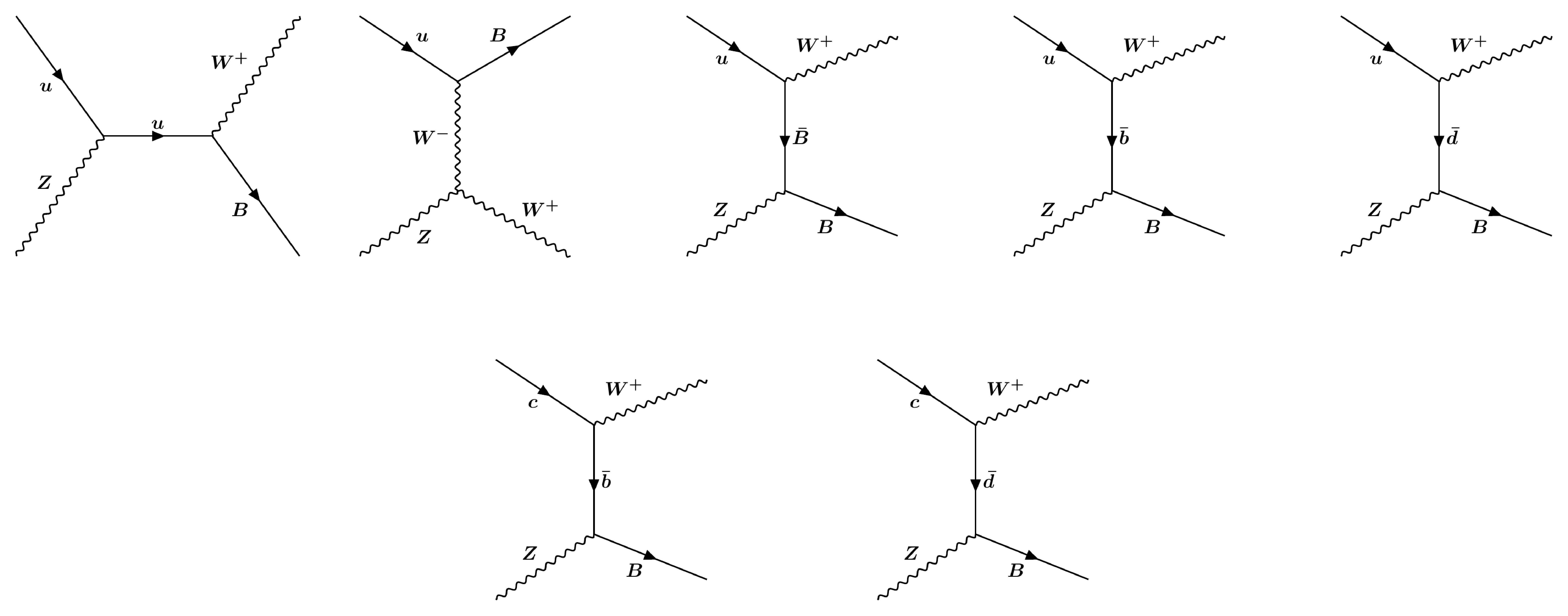}
    \caption{Representative tree-level Feynman diagrams for the $Z$-initiated subprocess $Z(u,c)\to W^{+}B$.}
    \label{fig:z_diagrams}
\end{figure*}

In the $Z$-initiated channel, the vector-like $B$ quark is produced through the interaction of an effective $Z$ boson radiated from the incoming muon with a parton from the proton. The relevant light-quark-initiated subprocesses can be written as
\begin{equation}
Zq\to W^+B,\qquad q=u,c.
\end{equation}
The up-quark-initiated contribution is dominant, while the charm-quark-initiated contribution remains numerically very small. Representative tree-level Feynman diagrams for the $Z$-initiated production mechanism are shown in Fig.~\ref{fig:z_diagrams}.

At the muon--proton collision level, the signal process can be written as
\begin{equation}
\mu p \rightarrow \mu W^{+} B X
\rightarrow \mu W^{+}(Zb)X
\rightarrow \mu \ell^{+}\ell^{-}bjjX,
\end{equation}
where $\ell=e,\mu$ and $X$ denotes the proton remnant. As in the photon-induced channel, the vector-like $B$ quark is reconstructed through the $B\to Zb$ decay mode, with $Z\to\ell^{+}\ell^{-}$ and $W\to jj$. The surviving muon remains associated with the $Z$-emission leg and is expected predominantly in the forward region.

Accordingly, the total cross section for $Z$-initiated single-$B$ production in muon--proton collisions can be expressed as
\begin{multline}
\sigma(\mu p\to \mu W^{+}BX)=
\sum_{q=u,c}\int dx_Z\int dx_q\,
f_{Z/\mu}^{\rm EVA}(x_Z) \\
\times f_{q/p}(x_q,\mu_F)\,
\hat{\sigma}(Zq\to W^{+}B),
\label{eq:sigma_mup_ZWB}
\end{multline}

where $f_{Z/\mu}^{\rm EVA}(x_Z)$ denotes the effective $Z$-boson distribution of the incoming muon, $f_{q/p}(x_q,\mu_F)$ is the corresponding quark distribution in the proton, and $\hat{\sigma}(Zq\to W^{+}B)$ represents the partonic cross section. In the adopted model, the relevant initial-state quark contributions arise from $q=u$ and $c$. The production rate is dominated by the up-quark-initiated subprocess, while the charm-quark contribution is numerically very small.

\subsection{Production Cross Section}

\begin{figure}[t]
    \centering
    \includegraphics[width=\linewidth]{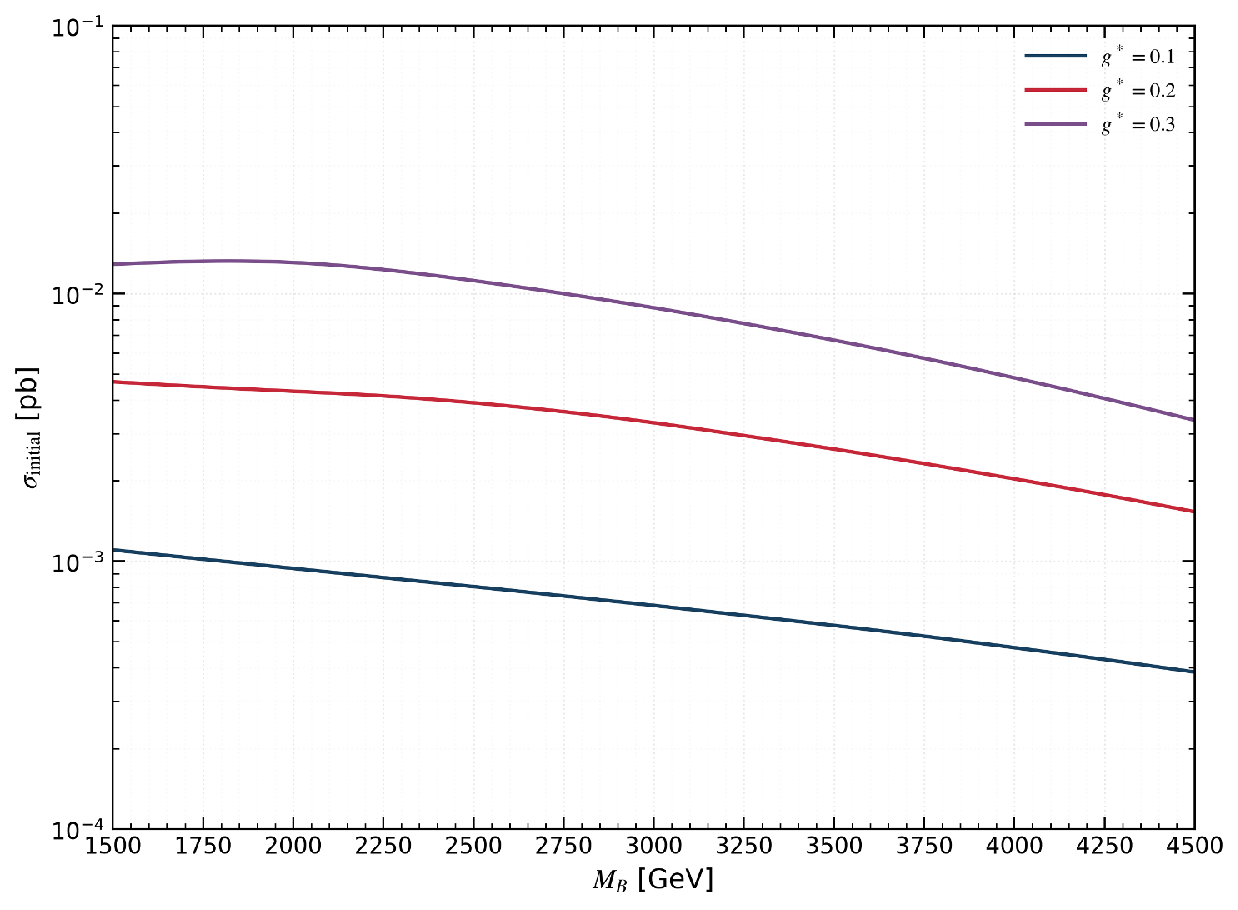}
    \caption{Initial cross section of the $Z$-initiated signal process, including the decay chain $B\to Zb$, $Z\to \ell^+\ell^-$, and $W\to jj$, as a function of the vector-like $B$-quark mass for different values of the coupling parameter $g^\ast$ at $R_L=0.05$.}
    \label{fig:Z_sigma_vs_MB}
\end{figure}

The initial cross section of the $Z$-initiated signal process, including the decay chain $B\to Zb$, $Z\to \ell^+\ell^-$, and $W\to jj$, was calculated as a function of the vector-like $B$-quark mass $M_B$ for different values of the coupling parameter $g^\ast$ at $R_L=0.05$. The results are presented in Fig.~\ref{fig:Z_sigma_vs_MB}. As in the photon-induced channel, the signal cross section decreases with increasing $M_B$, while larger values of $g^\ast$ enhance the signal rate.

\begin{figure}[t]
    \centering
    \includegraphics[width=0.48\textwidth]{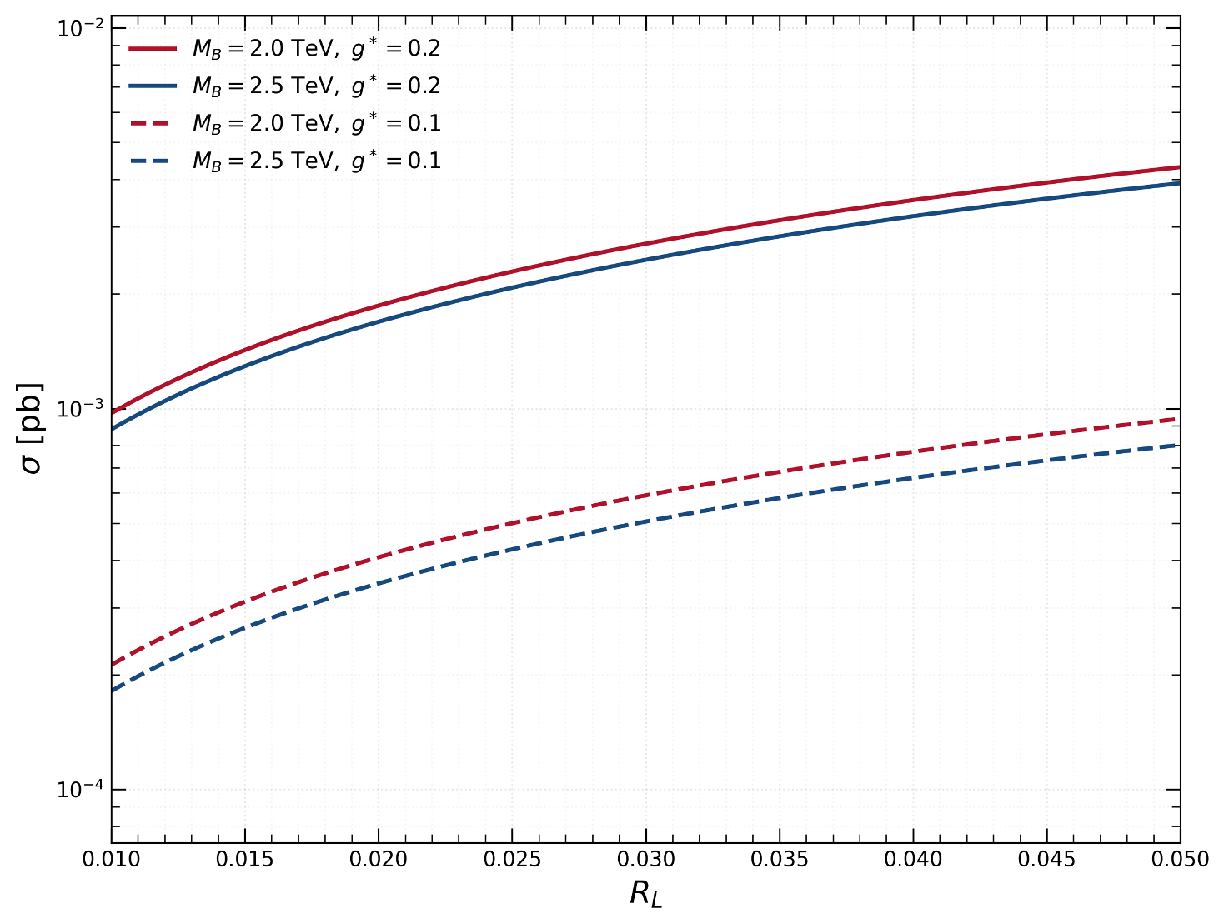}
    \caption{$Z$-initiated signal production cross section as a function of $R_L$ for $M_B=2.0$ and $2.5~\mathrm{TeV}$, shown for the benchmark coupling values $g^\ast=0.1$ and $0.2$.}
    \label{fig:Z_sigma_RL}
\end{figure}

The $Z$-initiated signal rate also depends on the mixing parameter $R_L$. This dependence is illustrated in Fig.~\ref{fig:Z_sigma_RL} for representative values of $M_B$ and $g^\ast$. Compared with the photon-induced production mechanism, the $Z$-initiated process contains additional tree-level contributions associated with the neutral-current interactions of the vector-like $B$ quark. In the adopted singlet-$B$ model, the relevant diagrams involve both the charged-current $BuW$ interaction and the neutral-current $BZd$ and $BZb$ vertices. Consequently, the $Z$-initiated production mechanism contains a richer diagrammatic structure than the photon-induced channel.

Since these interactions carry different first- and third-generation weights through $R_L$, and the corresponding partonic contributions are weighted by the proton parton distributions, the total $Z$-initiated signal rate exhibits a non-trivial dependence on the mixing parameter. In the considered parameter range, the dedicated $R_L$ scan shows that the signal cross section increases with increasing $R_L$.

As $g^\ast$ increases, the total decay width of the vector-like $B$ quark also becomes larger. The corresponding $\Gamma_B/M_B$ values are therefore taken into account when presenting the sensitivity regions.

\subsection{SM Backgrounds}

As in the photon-induced analysis, SM processes that can reproduce the same or a closely related reconstructed final-state topology are considered. The $WZj$, $WZb$, and $ZZj$ channels constitute the common background set for the two production mechanisms. For the $Z$-initiated analysis, the $Zjjj$ and $Zb\bar{b}j$ processes are additionally included in order to account for multijet and heavy-flavour contributions. The complete set of background processes is therefore

\begin{equation}
\begin{aligned}
Zp &\to WZj, \qquad
Zp \to WZb, \qquad
Zp \to ZZj, \\
Zp &\to Zjjj, \qquad
Zp \to Zb\bar{b}j .
\end{aligned}
\end{equation}

The corresponding samples were generated at parton level using the following commands
\begin{lstlisting}[
  breaklines=true,
  breakatwhitespace=true,
  basicstyle=\ttfamily\footnotesize
]
generate z p > w+ z j, (w+ > j j), (z > l+ l-)
generate z p > w+ z b, (w+ > j j), (z > l+ l-)
generate z p > z z j, (z > l+ l-), (z > j j)
generate z p > z j j j, (z > l+ l-)
generate z p > z b b~ j, (z > l+ l-)
\end{lstlisting}
for the $WZj$, $WZb$, $ZZj$, $Zjjj$, and $Zb\bar{b}j$ backgrounds, respectively. The same EVA beam configuration and proton PDF setup adopted for the $Z$-initiated signal generation were also used for these background samples. No analysis-level selection cuts were imposed at the event-generation stage. A minimal generation-level requirement of $p_T>50~\mathrm{GeV}$ was applied to the relevant jet or $b$-quark parton only where necessary to regulate the low-$p_T$ phase-space region and ensure stable numerical integration. This technical requirement is not part of the subsequent event-selection strategy.

The additional $Zjjj$ and $Zb\bar{b}j$ channels can also contribute to the reconstructed $bjj\ell^+\ell^-$ topology. The $Zjjj$ process represents a reducible multijet background in which a light jet is misidentified as a $b$ jet, whereas the $Zb\bar{b}j$ process provides a heavy-flavour background containing genuine $b$ jets. Their inclusion therefore provides a more complete description of the SM background contributions in the $Z$-initiated channel.

Accordingly, the $WZj$, $ZZj$, and $Zjjj$ processes are treated as reducible backgrounds, and a light-jet-to-$b$ misidentification probability of
\begin{equation}
\epsilon_{j\to b}=0.01
\end{equation}
is included in their normalization. In contrast, the $WZb$ and $Zb\bar{b}j$ processes contain genuine $b$ jets and are treated as heavy-flavour background contributions.

\subsection{Event Selection and Detector Smearing}

\begin{figure*}[t]
\centering

\begin{minipage}{0.48\textwidth}
    \centering
    \includegraphics[width=\textwidth]{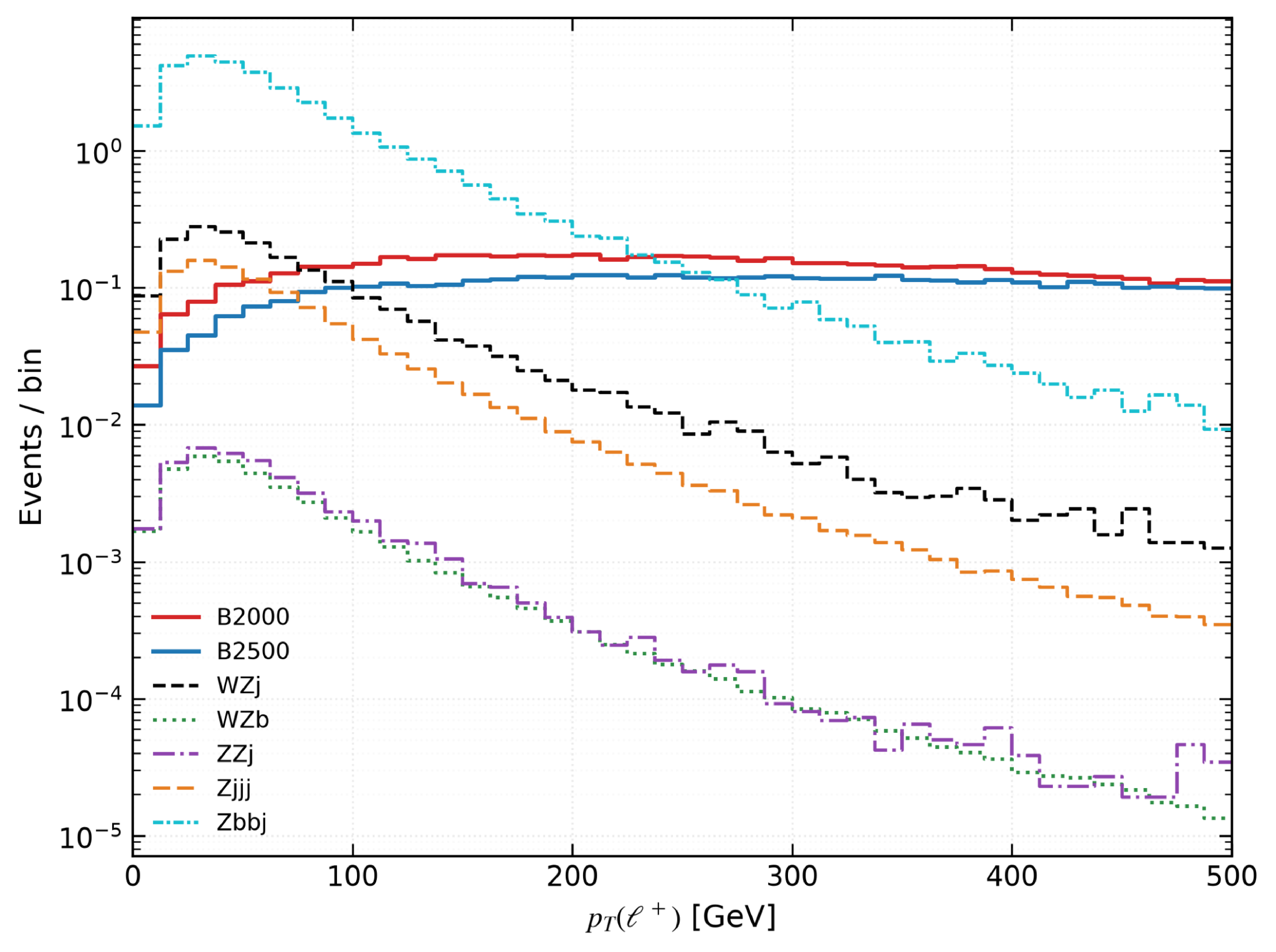}
    \par\small (a) $p_T(\ell)$ distribution
\end{minipage}
\hfill
\begin{minipage}{0.48\textwidth}
    \centering
    \includegraphics[width=\textwidth]{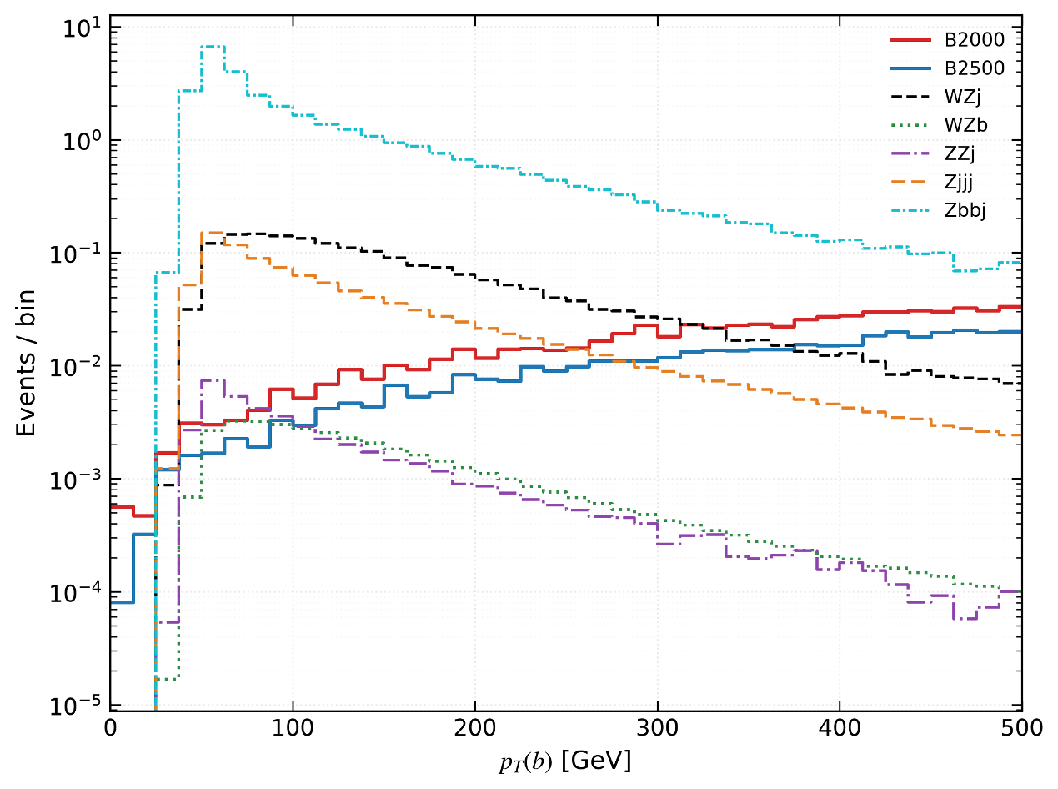}
    \par\small (b) $p_T(b)$ distribution
\end{minipage}

\vspace{0.4cm}

\begin{minipage}{0.48\textwidth}
    \centering
    \includegraphics[width=\textwidth]{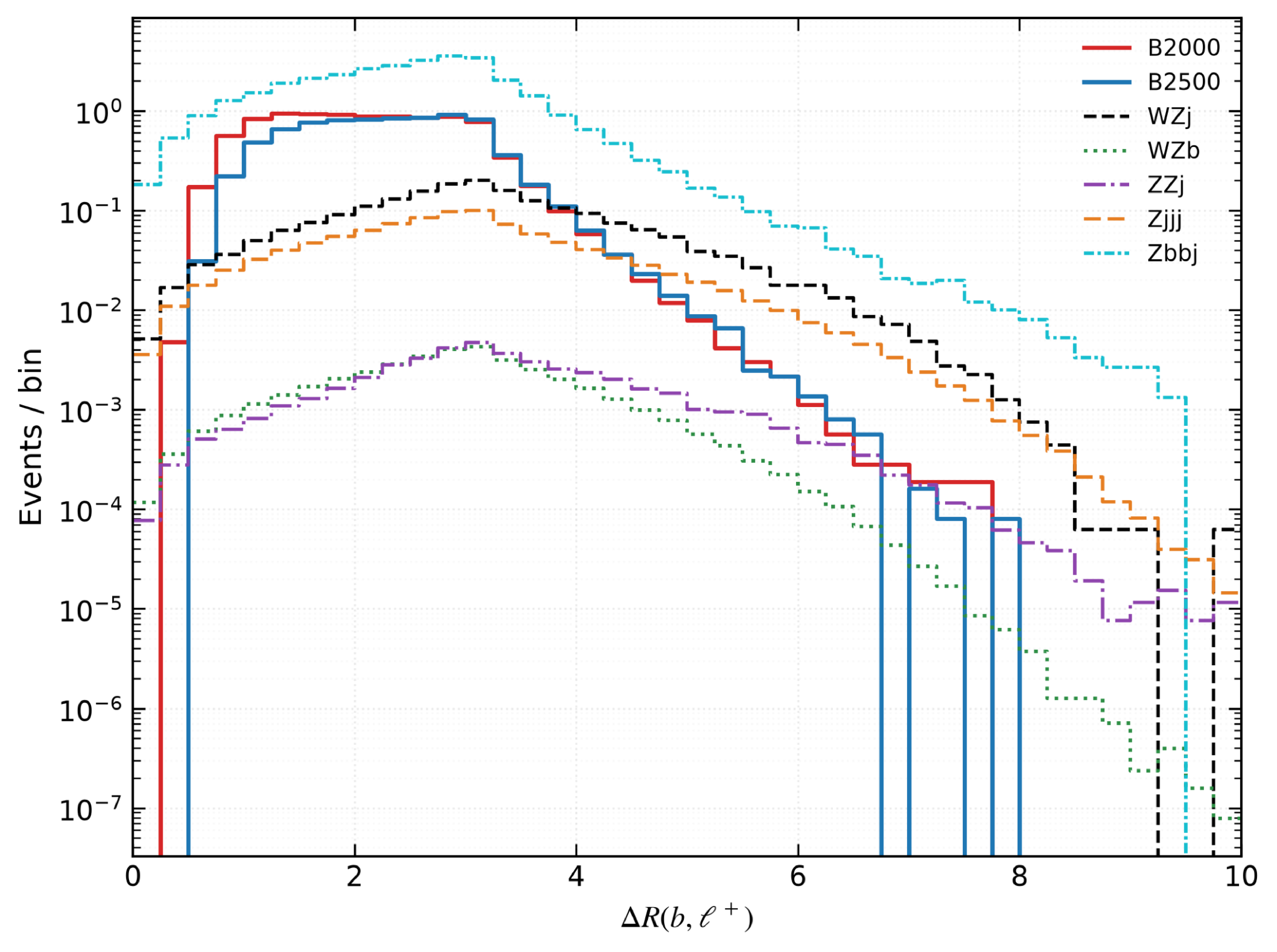}
    \par\small (c) $\Delta R(b,\ell)$ distribution
\end{minipage}
\hfill
\begin{minipage}{0.48\textwidth}
    \centering
    \includegraphics[width=\textwidth]{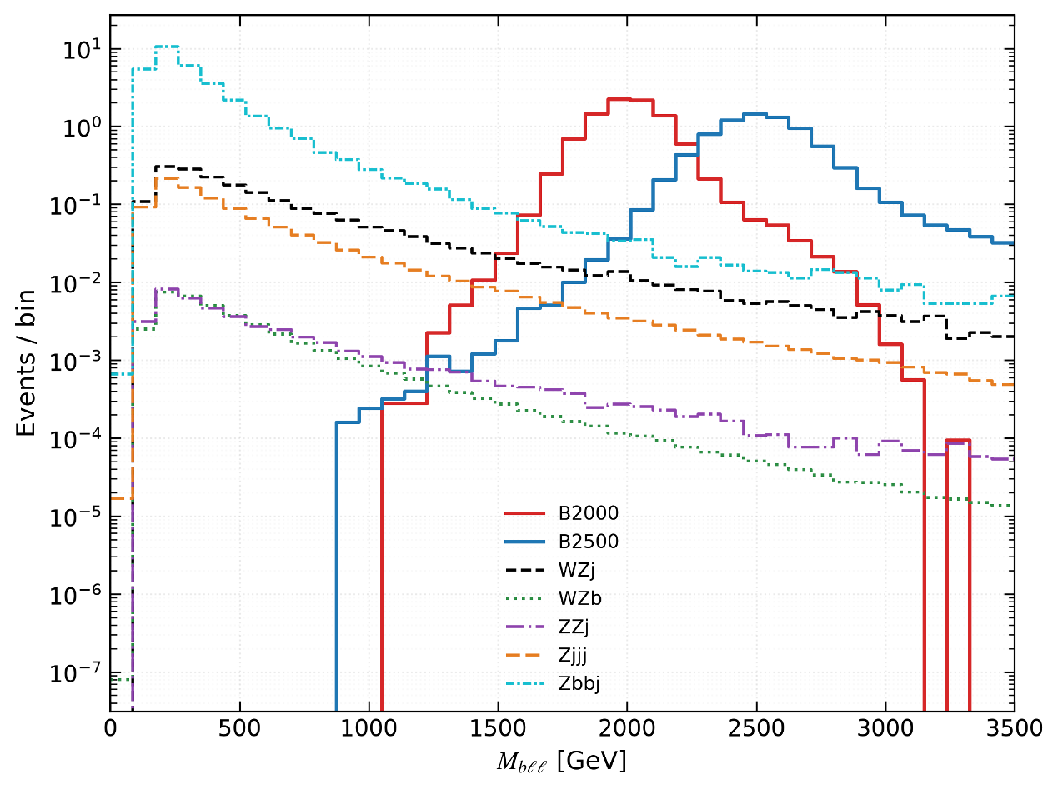}
    \par\small (d) $M_{b\ell\ell}$ distribution
\end{minipage}

\caption{Initial smeared kinematic distributions of the variables used in the
$Z$-initiated analysis for the signal and background processes. The signal
distributions are shown for the benchmark coupling value $g^\ast=0.1$.
The event yields are normalised to an integrated luminosity of
$\mathcal{L}=10~\mathrm{fb^{-1}}$.}
\label{fig:Z_kinematic_overview}
\end{figure*}

The same detector-smearing procedure adopted in the photon-induced analysis was applied to the $Z$-initiated signal and background samples. The corresponding smeared kinematic distributions are shown in Fig.~\ref{fig:Z_kinematic_overview}. Based on the different kinematic behaviour of the signal and background processes, the following selection requirements were applied sequentially.

As shown in Fig.~\ref{fig:Z_kinematic_overview}(a) and (b), the leptons and $b$ jets associated with the heavy $B$-quark decay tend to populate higher transverse-momentum regions than the background processes. The $b$-jet transverse-momentum distribution therefore provides useful signal-background discrimination. In the $Z$-initiated analysis, the requirement $p_T^b>50~\mathrm{GeV}$ is adopted consistently with the transverse-momentum threshold used in the generation of the relevant background samples. For leptons and light jets, the basic selection requirement $p_T>20~\mathrm{GeV}$ is applied.

Figure~\ref{fig:Z_kinematic_overview}(c) shows that the $Z$-initiated signal is mainly concentrated in an intermediate region of $\Delta R(b,\ell^+)$, whereas the background processes extend over a broader range. This behaviour indicates that the angular-separation variable provides useful discrimination between the signal and background processes. In contrast to the photon-induced channel, the signal populates a smaller $\Delta R(b,\ell^+)$ region, reflecting the different production kinematics of the two mechanisms.

Finally, Fig.~\ref{fig:Z_kinematic_overview}(d) shows the reconstructed heavy-quark invariant mass $M_{b\ell\ell}$. The signal exhibits a characteristic structure associated with the heavy $B$ resonance, whereas the background distributions are broader and smoother. These features provide an additional distinction between the signal and background processes.

Considering these kinematic features, selection cuts were sequentially applied to strengthen the signal-background separation as follows.

\textbf{Cut--1:} For detector acceptance and basic object selection,
\begin{itemize}
    \item $p_T^\ell>20~\mathrm{GeV}, \qquad |\eta^\ell|<2.5,$
    \item $p_T^b>50~\mathrm{GeV}, \qquad |\eta^b|<4.0,$
    \item $p_T^j>20~\mathrm{GeV}, \qquad |\eta^j|<5.0.$
\end{itemize}

\textbf{Cut--2 (Angular Separation):}
\[
1.0 < \Delta R(b,\ell^+) < 3.5.
\]

\textbf{Cut--3 (Hadronic Boson Reconstruction):}
\[
70~\mathrm{GeV}<M_{jj}<90~\mathrm{GeV}.
\]

\textbf{Cut--4 (Leptonic $Z$ Reconstruction):}
\[
80~\mathrm{GeV}<M_{\ell\ell}<100~\mathrm{GeV}.
\]

\begin{table*}[t]
\centering
\caption{Cut-flow cross sections for the $Z$-initiated signal and background processes after successive selection criteria. The signal results are shown for the benchmark coupling $g^\ast=0.1$ and two representative VLQ masses. Detector-level energy smearing is applied throughout the cut-flow. All cross sections are given in pb.}
\label{tab:cutflow_z_smear}
\resizebox{\textwidth}{!}{%
\begin{tabular}{lccccccc}
\hline
Selection
& Signal ($M_B=2000~\mathrm{GeV}$)
& Signal ($M_B=2500~\mathrm{GeV}$)
& $WZj$
& $WZb$
& $ZZj$
& $Zjjj$
& $Zb\bar{b}j$ \\
\hline

Initial
& $9.399\times10^{-4}$
& $8.040\times10^{-4}$
& $2.000\times10^{-4}$
& $3.954\times10^{-6}$
& $4.562\times10^{-6}$
& $1.038\times10^{-4}$
& $3.316\times10^{-3}$ \\

Cut 1
& $8.162\times10^{-4}$
& $7.154\times10^{-4}$
& $7.900\times10^{-5}$
& $1.900\times10^{-6}$
& $1.934\times10^{-6}$
& $5.639\times10^{-5}$
& $1.861\times10^{-3}$ \\

Cut 2
& $7.165\times10^{-4}$
& $6.588\times10^{-4}$
& $5.192\times10^{-5}$
& $1.331\times10^{-6}$
& $1.213\times10^{-6}$
& $3.923\times10^{-5}$
& $1.469\times10^{-3}$ \\

Cut 3
& $6.200\times10^{-4}$
& $5.694\times10^{-4}$
& $4.450\times10^{-5}$
& $1.146\times10^{-6}$
& $5.076\times10^{-7}$
& $2.482\times10^{-6}$
& $4.457\times10^{-5}$ \\

Cut 4
& $5.845\times10^{-4}$
& $5.362\times10^{-4}$
& $4.193\times10^{-5}$
& $1.077\times10^{-6}$
& $4.785\times10^{-7}$
& $2.343\times10^{-6}$
& $4.152\times10^{-5}$ \\

Efficiency (\%)
& $62.19$
& $66.70$
& $20.96$
& $27.24$
& $10.49$
& $2.26$
& $1.25$ \\

\hline
\end{tabular}%
}
\end{table*}

The quantitative effect of the applied selection criteria is summarized in Table~\ref{tab:cutflow_z_smear}. As seen from the table, the signal cross sections are reduced only moderately after the successive cuts. For the benchmark points $M_B=2.0$ and $2.5~\mathrm{TeV}$ at $g^\ast=0.1$, the final signal cross sections are $5.845\times10^{-4}$ pb and $5.362\times10^{-4}$ pb, corresponding to total efficiencies of $62.19\%$ and $66.70\%$, respectively. In contrast, the SM background contributions are more strongly affected by the selection requirements. In particular, the hadronic-boson reconstruction condition on $M_{jj}$ provides a substantial suppression of the multijet backgrounds, especially for the $Zjjj$ and $Zb\bar{b}j$ channels, while it retains a large fraction of the signal events. After all selection criteria, the $WZj$, $WZb$, $ZZj$, $Zjjj$, and $Zb\bar{b}j$ background cross sections are reduced to $4.193\times10^{-5}$ pb, $1.077\times10^{-6}$ pb, $4.785\times10^{-7}$ pb, $2.343\times10^{-6}$ pb, and $4.152\times10^{-5}$ pb, respectively. The light-jet mistag probability is included in the quoted $WZj$, $ZZj$, and $Zjjj$ rates, while no $b$-tagging efficiency is applied in the cut-flow table. These results demonstrate that the adopted conservative event-selection strategy provides a clear separation between the $Z$-initiated signal and the SM backgrounds while it preserves a large signal efficiency.

\section{Statistical Analysis and Sensitivity Results}
\label{sec:analysis}

\subsection{Statistical Framework}

To quantify the discovery and exclusion sensitivities, the statistical significance is evaluated using the Asimov approach. The signal event yield is defined as
\begin{equation}
s=\sigma_S\,\mathcal{L}\,\varepsilon_b,
\end{equation}
where $\sigma_S$ is the signal cross section after the final event selection, $\mathcal{L}$ is the integrated luminosity, and $\varepsilon_b=0.85$ is the tagging efficiency for a genuine $b$ jet.

For the photon-induced analysis, the expected background yield is
\begin{equation}
b_{\gamma}
=
\mathcal{L}
\left[
\sigma_{WZb}\,\varepsilon_b
+\sigma_{WZj}
+\sigma_{ZZj}
\right].
\label{eq:b_photon}
\end{equation}
For the $Z$-initiated analysis, the corresponding background yield is
\begin{equation}
b_Z
=
\mathcal{L}
\left[
\left(
\sigma_{WZb}
+\sigma_{Zb\bar{b}j}
\right)\varepsilon_b
+\sigma_{WZj}
+\sigma_{ZZj}
+\sigma_{Zjjj}
\right].
\label{eq:b_Z}
\end{equation}

The $b$-tagging efficiency is applied to the signal and to the background processes containing genuine $b$ jets. For the $WZj$, $ZZj$, and $Zjjj$ backgrounds, the $b$-tagged candidate originates from a misidentified light jet. A light-jet-to-$b$ mistag probability of $1\%$ is already included in the quoted cross sections of these reducible backgrounds. Therefore, no additional factor of $\varepsilon_b$ is applied to these contributions, avoiding double counting of the tagging factors~\cite{ATLASbtag2016,CMSbtag2017,FCCbtag2021}.

When the relative background systematic uncertainty is denoted by $\delta_B$, the discovery significance is evaluated using the Asimov expression~\cite{Cowan2011Asymptotic}:
\begin{multline}
Z_{\rm disc}
=
\Bigg\{
2 \Bigg[
(s+b)\,
\ln\!\left(
\frac{(s+b)\,(1+\delta_B^2 b)}
{b+\delta_B^2 b\,(s+b)}
\right)
\\
-\frac{1}{\delta_B^2}\,
\ln\!\left(
1+\frac{\delta_B^2 s}
{1+\delta_B^2 b}
\right)
\Bigg]
\Bigg\}^{1/2}.
\label{eq:Zdisc_delta}
\end{multline}

In the limit $\delta_B\to0$, this expression reduces to
\begin{equation}
Z_{\rm disc}
=
\sqrt{
2\left[
(s+b)\ln\left(1+\frac{s}{b}\right)-s
\right]
}.
\label{eq:Zdisc_delta0}
\end{equation}

The expected exclusion sensitivity is evaluated using the corresponding Asimov-based exclusion significance~\cite{Cowan2011Asymptotic,Read2002CLs}:
\begin{multline}
Z_{\rm excl}
=
\Bigg\{
2\left[
s-b\ln\!\left(\frac{b+s+x}{2b}\right)
\right]
-\frac{2}{\delta_B^2}
\ln\!\left(\frac{b-s+x}{2b}\right)
\\
-(b+s-x)
\left(
1+\frac{1}{\delta_B^2 b}
\right)
\Bigg\}^{1/2},
\label{eq:Zexcl_delta}
\end{multline}
where
\begin{equation}
x=
\sqrt{
(s+b)^2
-
\frac{4\,\delta_B^2\,s\,b^2}
{1+\delta_B^2 b}
}.
\label{eq:x_def}
\end{equation}

In the limit $\delta_B\to0$, the exclusion significance reduces to
\begin{equation}
Z_{\rm excl}
=
\sqrt{
2\left[
s-b\ln\left(1+\frac{s}{b}\right)
\right]
}.
\label{eq:Zexcl_delta0}
\end{equation}

In this study, a discovery is defined by $Z_{\rm disc}\geq5$, corresponding to a $5\sigma$ significance. The expected $95\%$ C.L. exclusion sensitivity is defined by $Z_{\rm excl}\geq1.65$, corresponding approximately to a one-sided Gaussian confidence level.

\subsection{Photon-Induced Sensitivity}

\begin{figure*}[t]
\centering

\begin{minipage}{0.48\textwidth}
    \centering
    \includegraphics[width=\textwidth]{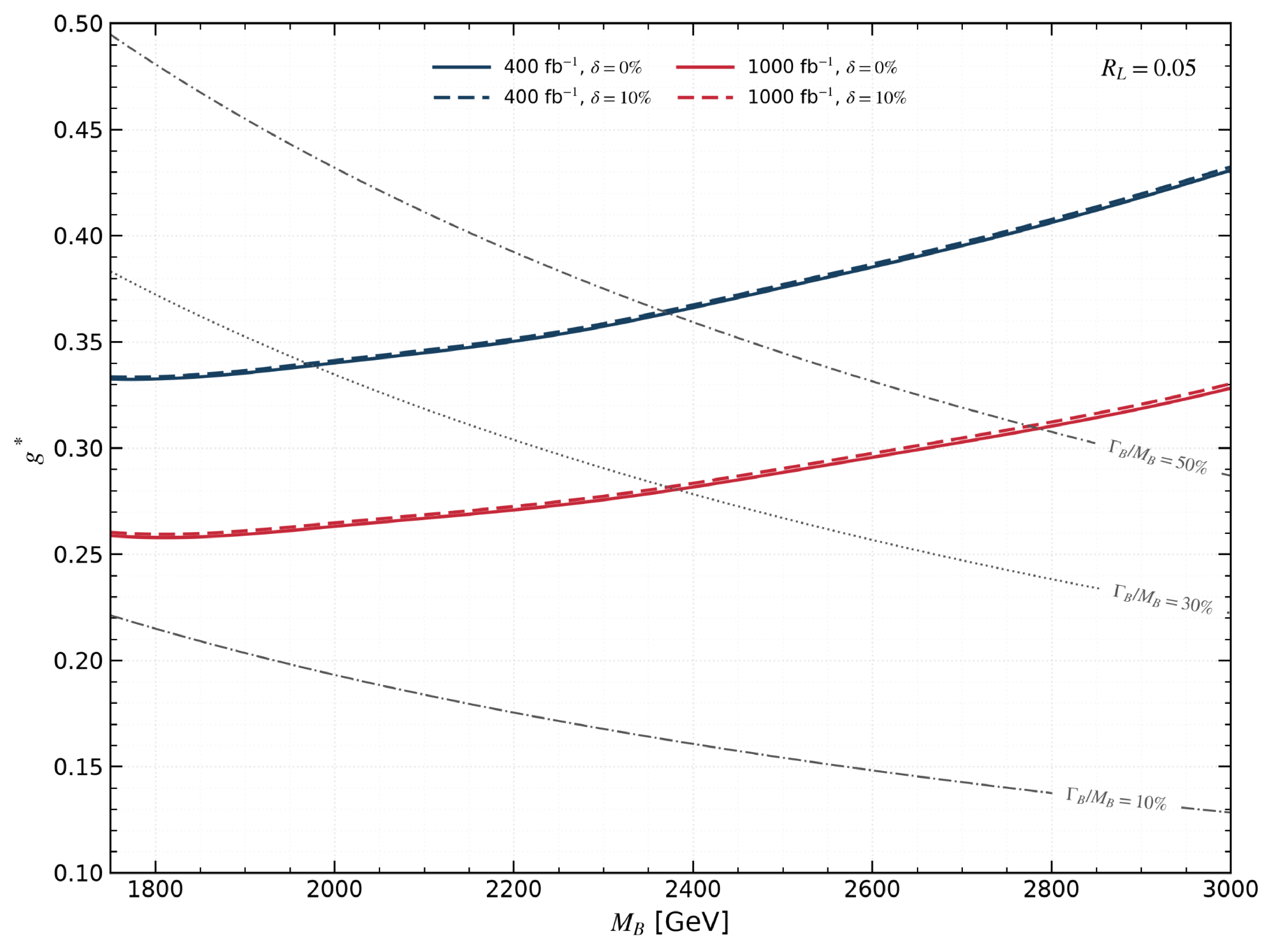}
    \par\small (a) Expected $5\sigma$ discovery reach
\end{minipage}
\hfill
\begin{minipage}{0.48\textwidth}
    \centering
    \includegraphics[width=\textwidth]{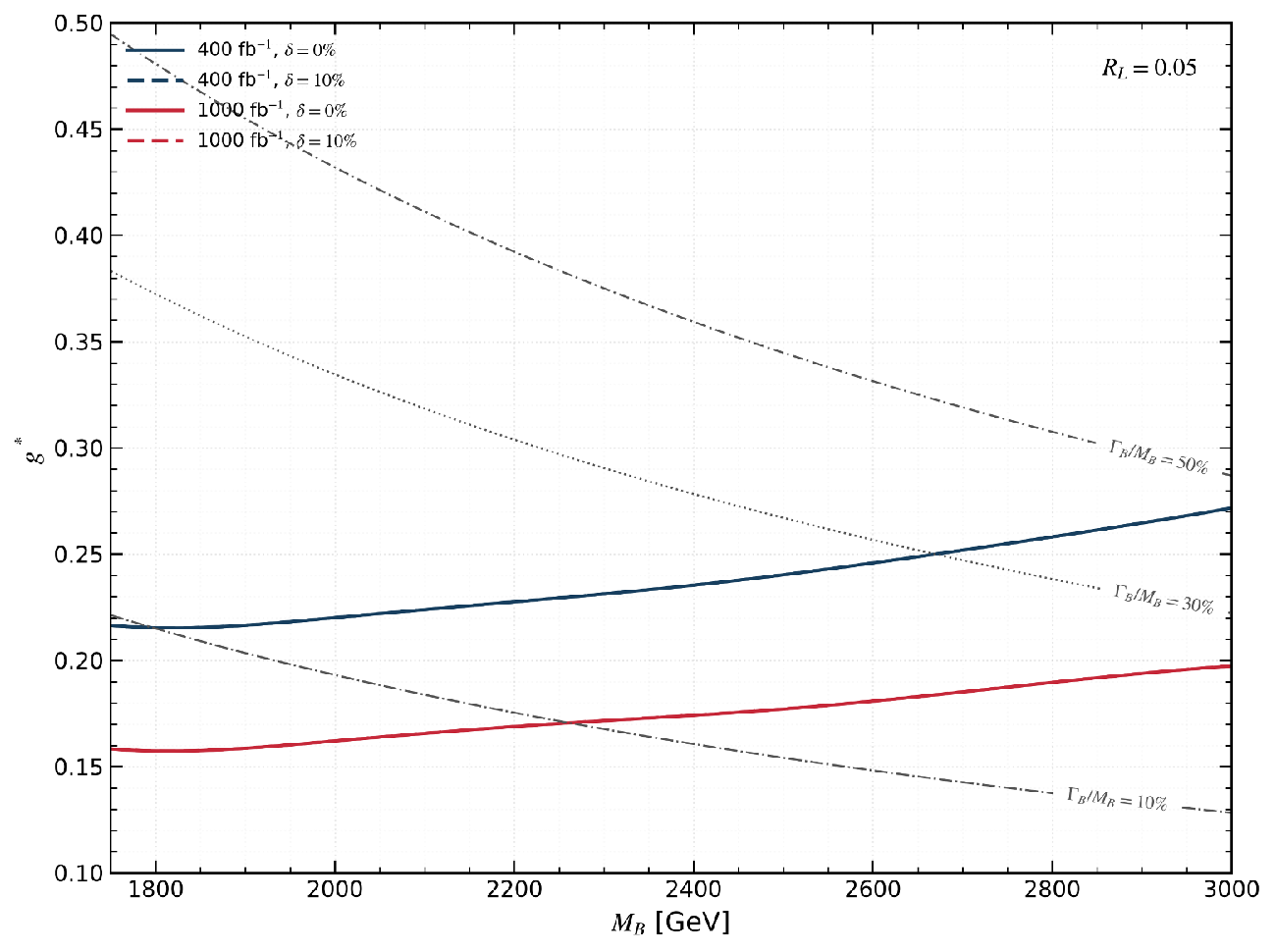}
    \par\small (b) Expected $95\%$ C.L. exclusion reach
\end{minipage}

\caption{Expected sensitivity to the photon-induced production channel
in the $g^\ast$--$M_B$ plane for $R_L=0.05$.
The discovery and exclusion reaches are shown for integrated luminosities
of $\mathcal{L}=400$ and $1000~\mathrm{fb^{-1}}$, considering
background systematic uncertainties of $\delta_B=0$ and $\delta_B=10\%$.}
\label{fig:photon_RL005_sensitivity}

\end{figure*}

\begin{figure*}[t]
\centering

\begin{minipage}{0.48\textwidth}
    \centering
    \includegraphics[width=\textwidth]{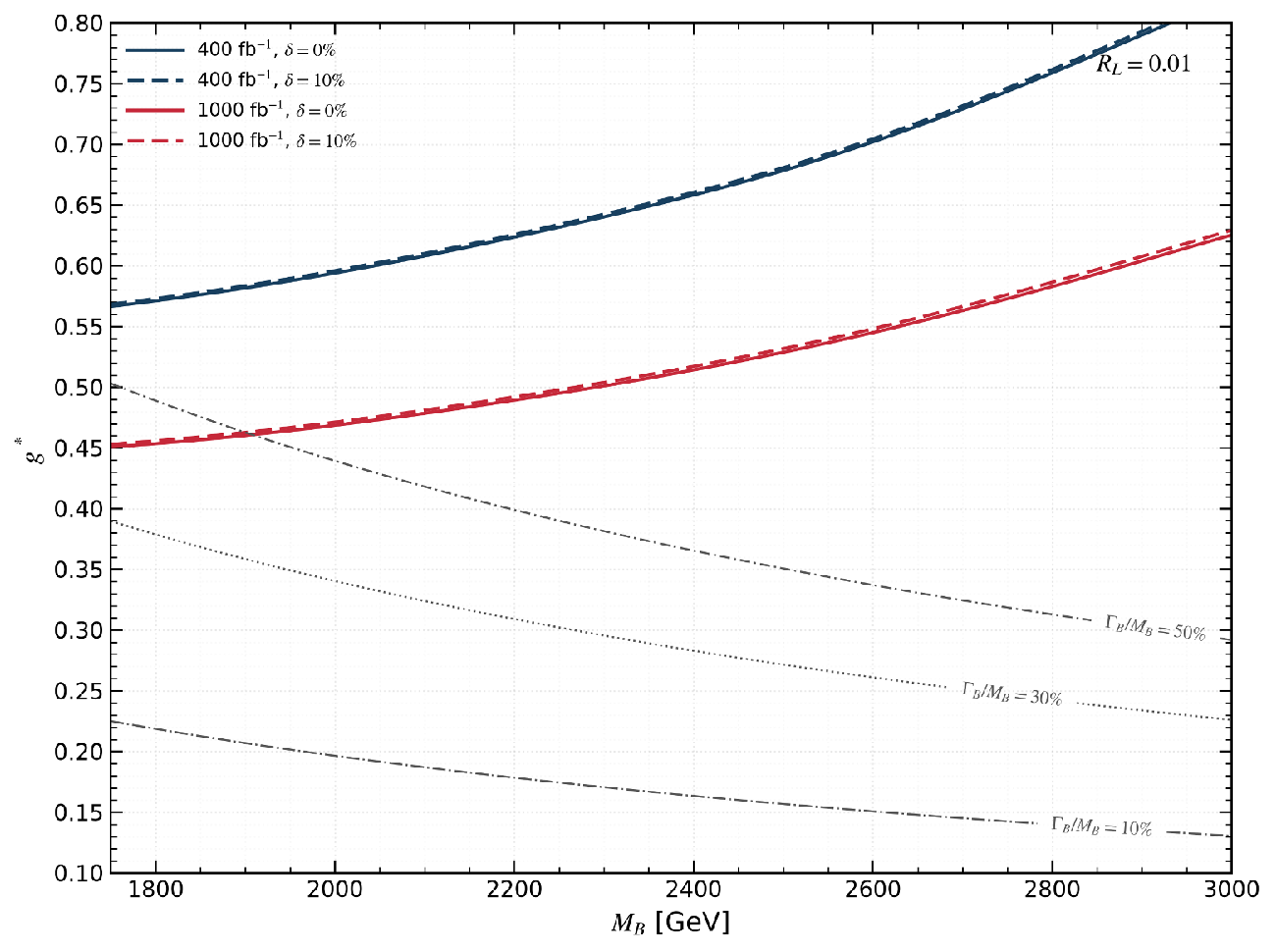}
    \par\small (a) Expected $5\sigma$ discovery reach
\end{minipage}
\hfill
\begin{minipage}{0.48\textwidth}
    \centering
    \includegraphics[width=\textwidth]{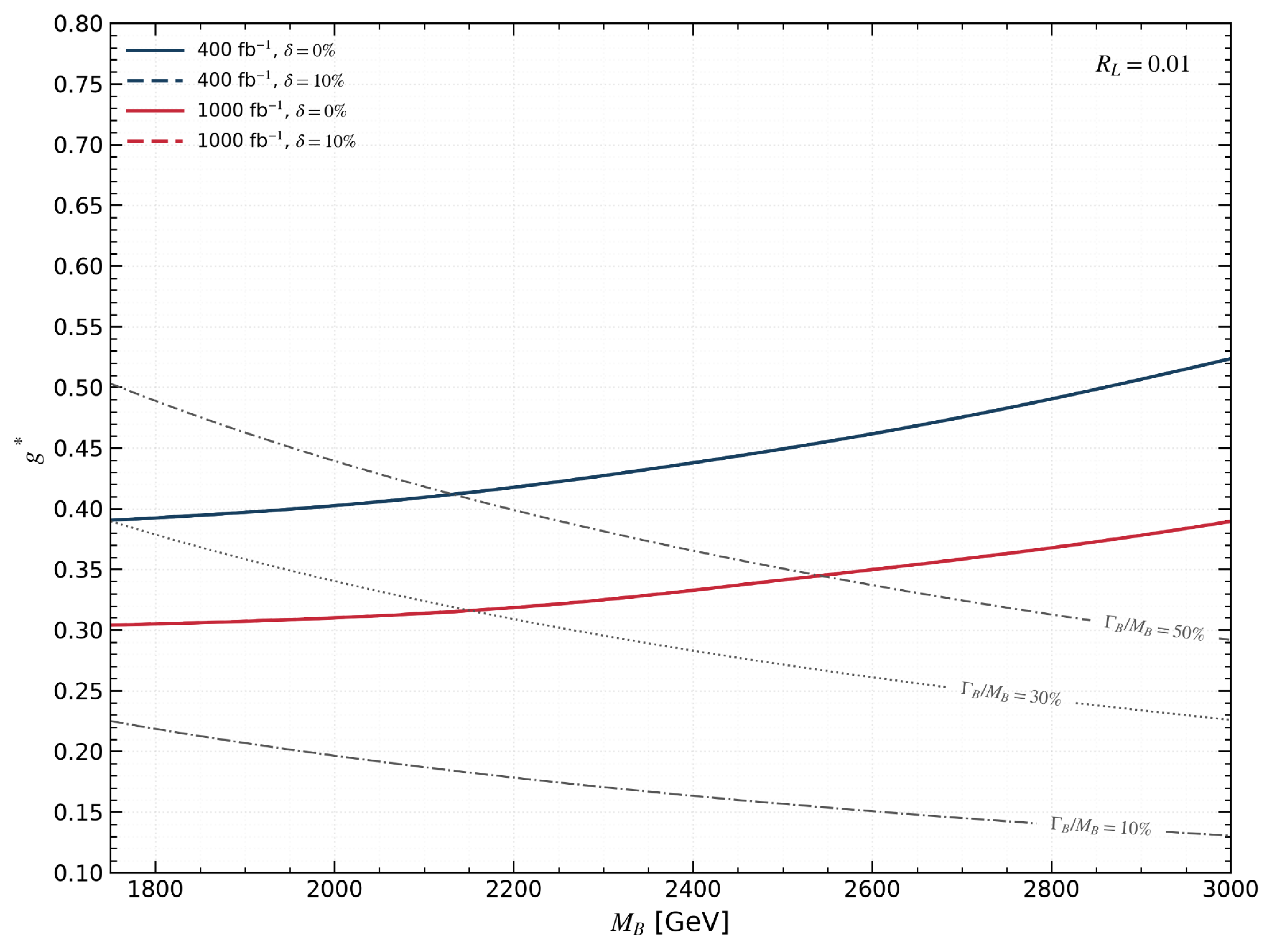}
    \par\small (b) Expected $95\%$ C.L. exclusion reach
\end{minipage}

\caption{Expected sensitivity to the photon-induced production channel
in the $g^\ast$--$M_B$ plane for $R_L=0.01$.
The discovery and exclusion reaches are shown for integrated luminosities
of $\mathcal{L}=400$ and $1000~\mathrm{fb^{-1}}$, considering
background systematic uncertainties of $\delta_B=0$ and $\delta_B=10\%$.}
\label{fig:photon_RL001_sensitivity}

\end{figure*}

Using the statistical framework described above, the discovery and exclusion sensitivities of the photon-induced channel were evaluated in the $g^\ast$--$M_B$ parameter space. The results were obtained for integrated luminosities of $\mathcal{L}=400$ and $1000~\mathrm{fb^{-1}}$, considering background systematic uncertainties of $\delta=0$ and $\delta=10\%$. A $b$-tagging efficiency of $\varepsilon_b=0.85$ was included in the event yields. The sensitivities are presented for the two benchmark mixing values $R_L=0.05$ and $R_L=0.01$ in Figs.~\ref{fig:photon_RL005_sensitivity} and~\ref{fig:photon_RL001_sensitivity}, respectively.

For $R_L=0.05$, the expected $5\sigma$ discovery reach and the $95\%$ C.L. exclusion sensitivity are shown in Fig.~\ref{fig:photon_RL005_sensitivity}. As can be seen from Fig.~\ref{fig:photon_RL005_sensitivity}(a), the minimum coupling required for discovery increases with the VLQ mass $M_B$. Increasing the integrated luminosity from $400$ to $1000~\mathrm{fb^{-1}}$ provides a clear improvement in the accessible parameter region. For $\mathcal{L}=1000~\mathrm{fb^{-1}}$, the $5\sigma$ discovery reach is approximately $g^\ast\simeq0.26$--$0.33$ over the mass interval $M_B=2$--$3~\mathrm{TeV}$, while for $\mathcal{L}=400~\mathrm{fb^{-1}}$ the corresponding reach is approximately $g^\ast\simeq0.34$--$0.43$.

The corresponding $95\%$ C.L. exclusion sensitivity is presented in Fig.~\ref{fig:photon_RL005_sensitivity}(b). As expected, smaller values of $g^\ast$ can be probed in the exclusion analysis. For $\mathcal{L}=1000~\mathrm{fb^{-1}}$, the exclusion reach extends down to approximately $g^\ast\simeq0.16$--$0.20$ in the mass range $M_B=2$--$3~\mathrm{TeV}$, whereas for $\mathcal{L}=400~\mathrm{fb^{-1}}$ the corresponding sensitivity is approximately $g^\ast\simeq0.22$--$0.28$. 

The resonance-width regimes corresponding to the sensitivity contours are indicated by the ratios $\Gamma_B/M_B=10\%$, $30\%$, and $50\%$. These contours show that the exclusion sensitivity mainly extends into relatively narrower-width regions, while part of the $5\sigma$ discovery reach enters regions with larger $\Gamma_B/M_B$. Since the resonant decay-chain simulation employed here does not constitute a complete non-resonant off-shell and interference calculation, results extending into the broad-width region should be interpreted with appropriate caution.

The sensitivity for the smaller mixing value $R_L=0.01$ is shown in Fig.~\ref{fig:photon_RL001_sensitivity}. Compared with the $R_L=0.05$ benchmark, the smaller signal rate leads to a weaker discovery and exclusion sensitivity, so that larger values of $g^\ast$ are required for a given $M_B$. Nevertheless, increasing the integrated luminosity from $400$ to $1000~\mathrm{fb^{-1}}$ significantly improves the sensitivity and enlarges the accessible parameter region. The best sensitivity is therefore obtained for $\mathcal{L}=1000~\mathrm{fb^{-1}}$. The same width-to-mass contours are included to indicate the resonance-width regime associated with the obtained sensitivities. 

For both mixing scenarios, the difference between the $\delta=0$ and $\delta=10\%$ contours remain small. This behaviour follows from the strong suppression of the SM backgrounds after the final event selection. Consequently, the photon-induced sensitivity is mainly controlled by the remaining signal event yield, while the assumed background systematic uncertainty produces only a minor modification of the discovery and exclusion reaches.

\subsection{$Z$-Initiated Sensitivity}

Using the statistical framework described above, the discovery and exclusion sensitivities of the $Z$-initiated channel were evaluated in the $g^\ast$--$M_B$ parameter space. The results are presented for the benchmark mixing values $R_L=0.05$ and $R_L=0.01$, considering integrated luminosities of $\mathcal{L}=100$ and $500~\mathrm{fb^{-1}}$, background systematic uncertainties of $\delta_B=0$ and $\delta_B=10\%$, and a $b$-tagging efficiency of $\varepsilon_b=0.85$.

\begin{figure*}[t]
\centering

\begin{minipage}{0.48\textwidth}
    \centering
    \includegraphics[width=\textwidth]{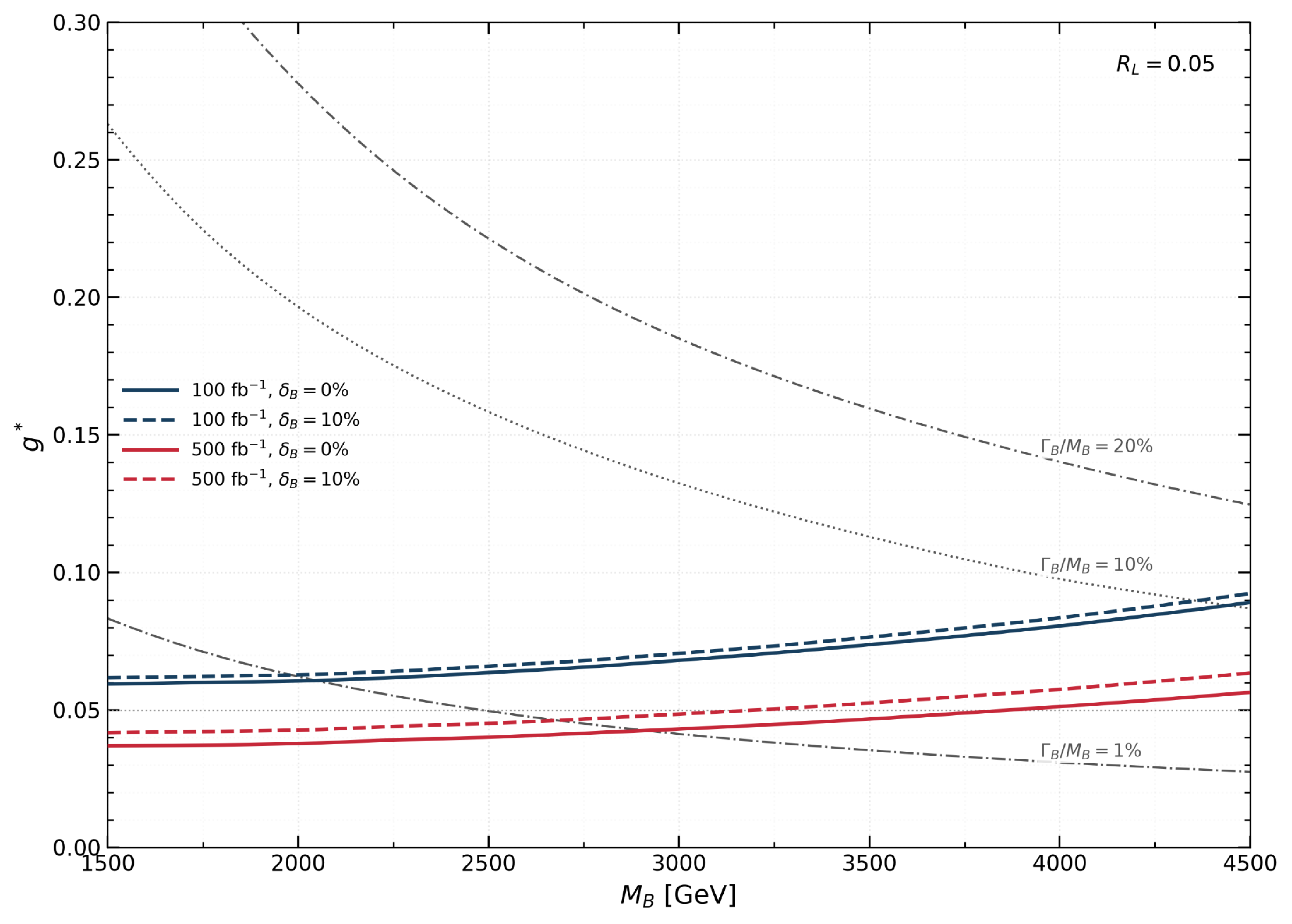}
    \par\small (a) Expected $5\sigma$ discovery reach
\end{minipage}
\hfill
\begin{minipage}{0.48\textwidth}
    \centering
    \includegraphics[width=\textwidth]{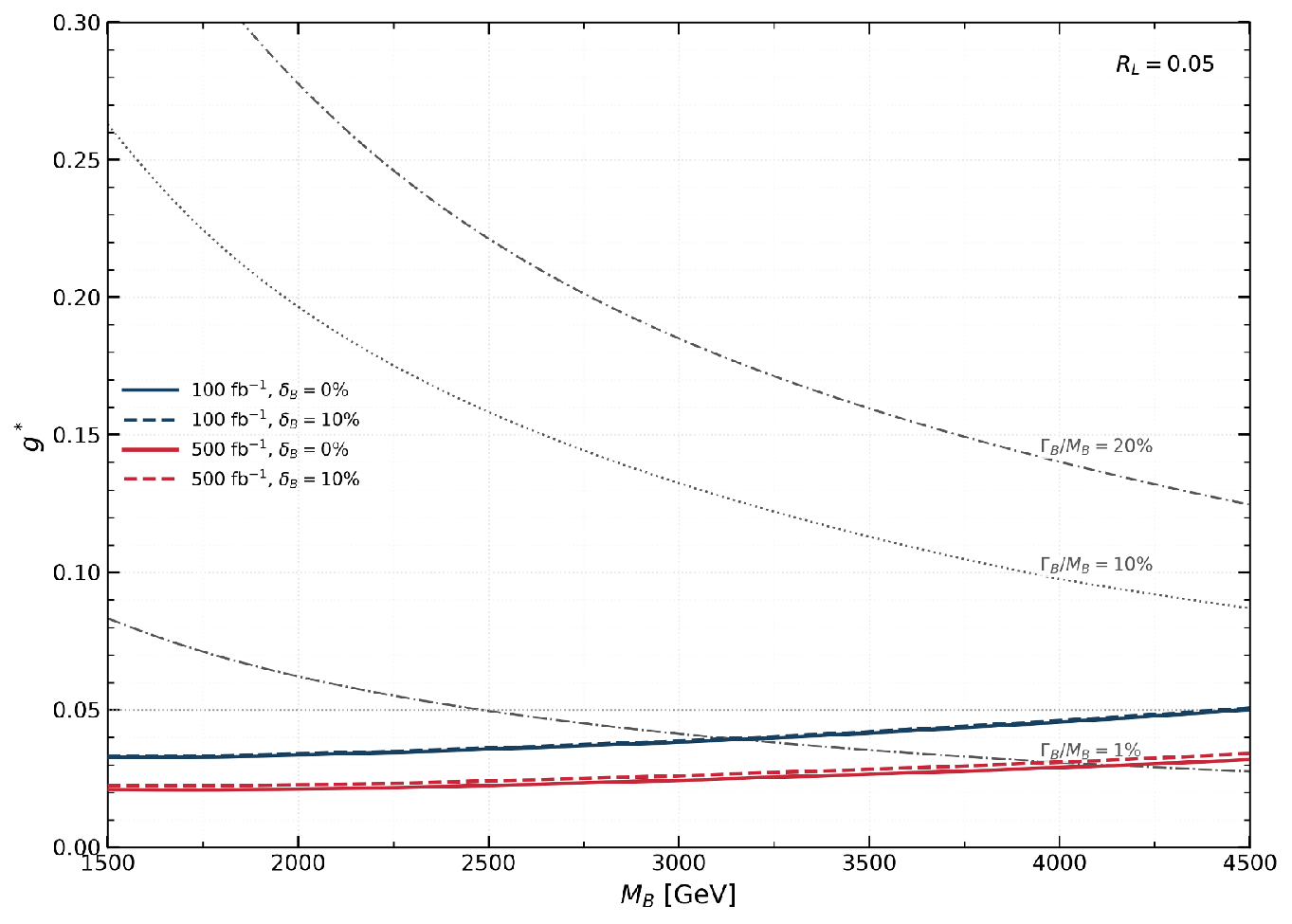}
    \par\small (b) Expected $95\%$ C.L. exclusion reach
\end{minipage}

\caption{Expected sensitivity to the $Z$-initiated production channel
in the $g^\ast$--$M_B$ plane for $R_L=0.05$.
The discovery and exclusion reaches are shown for integrated luminosities
of $\mathcal{L}=100$ and $500~\mathrm{fb^{-1}}$, considering
background systematic uncertainties of $\delta_B=0$ and
$\delta_B=10\%$.}
\label{fig:Z_RL005_sensitivity}

\end{figure*}

For $R_L=0.05$, the expected $5\sigma$ discovery and $95\%$ C.L. exclusion sensitivities are shown in Fig.~\ref{fig:Z_RL005_sensitivity}. Over the mass interval $M_B=2$--$4.5~\mathrm{TeV}$, the discovery reach varies approximately over $g^\ast\simeq0.061$--$0.089$ for $\mathcal{L}=100~\mathrm{fb^{-1}}$. Increasing the integrated luminosity to $\mathcal{L}=500~\mathrm{fb^{-1}}$ improves the corresponding sensitivity to approximately $g^\ast\simeq0.038$--$0.056$. The $95\%$ C.L. exclusion reach extends to approximately $g^\ast\simeq0.034$--$0.050$ and $g^\ast\simeq0.021$--$0.032$ for $\mathcal{L}=100$ and $500~\mathrm{fb^{-1}}$, respectively. The inclusion of a $10\%$ background systematic uncertainty produces a relatively small shift in the sensitivity contours.

\begin{figure*}[t]
\centering

\begin{minipage}{0.48\textwidth}
    \centering
    \includegraphics[width=\textwidth]{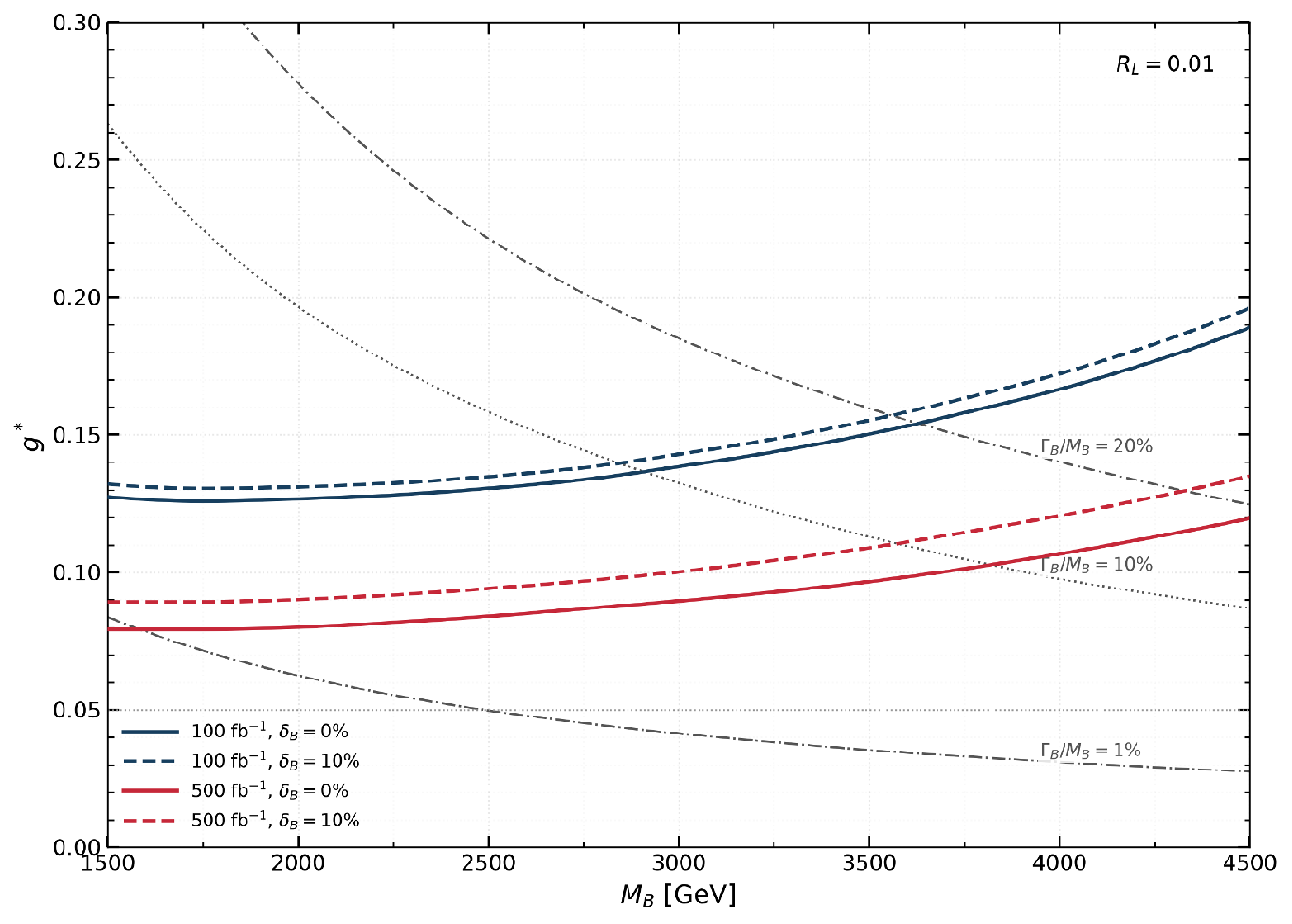}
    \par\small (a) Expected $5\sigma$ discovery reach
\end{minipage}
\hfill
\begin{minipage}{0.48\textwidth}
    \centering
    \includegraphics[width=\textwidth]{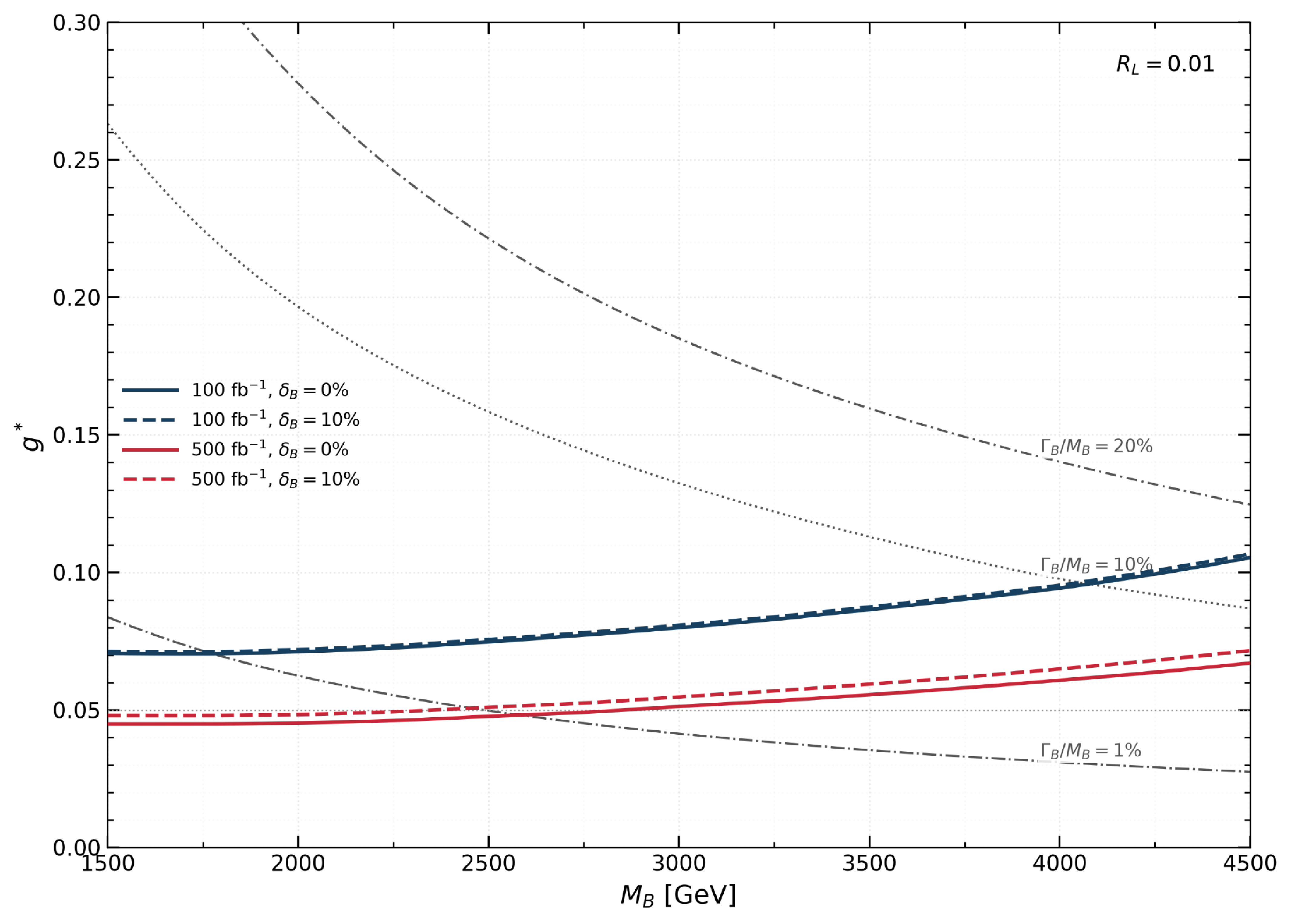}
    \par\small (b) Expected $95\%$ C.L. exclusion reach
\end{minipage}

\caption{Expected sensitivity to the $Z$-initiated production channel
in the $g^\ast$--$M_B$ plane for $R_L=0.01$.
The discovery and exclusion reaches are shown for integrated luminosities
of $\mathcal{L}=100$ and $500~\mathrm{fb^{-1}}$, considering
background systematic uncertainties of $\delta_B=0$ and
$\delta_B=10\%$.}
\label{fig:Z_RL001_sensitivity}

\end{figure*}

The sensitivity for the smaller mixing value $R_L=0.01$ is shown in Fig.~\ref{fig:Z_RL001_sensitivity}. As expected from the reduced signal rate, larger values of $g^\ast$ are required to obtain the same statistical significance. Over the mass interval $M_B=2$--$4.5~\mathrm{TeV}$, the $5\sigma$ discovery reach varies approximately over $g^\ast\simeq0.127$--$0.189$ for $\mathcal{L}=100~\mathrm{fb^{-1}}$ and $g^\ast\simeq0.080$--$0.120$ for $\mathcal{L}=500~\mathrm{fb^{-1}}$. The corresponding $95\%$ C.L. exclusion sensitivity extends over approximately $g^\ast\simeq0.071$--$0.105$ and $g^\ast\simeq0.045$--$0.067$ for $\mathcal{L}=100$ and $500~\mathrm{fb^{-1}}$, respectively. The effect of the $10\%$ background systematic uncertainty again produces only a relatively small modification of the sensitivity.

The resonance-width regimes associated with the sensitivity contours are indicated by $\Gamma_B/M_B=1\%$, $10\%$, and $20\%$. For $R_L=0.05$ and $\mathcal{L}=500~\mathrm{fb^{-1}}$, both the discovery and exclusion reaches remain below $\Gamma_B/M_B=10\%$ throughout the considered $M_B=2$--$4.5~\mathrm{TeV}$ mass interval. At $M_B=4.5~\mathrm{TeV}$ and $\delta_B=0$, the corresponding width-to-mass ratios are approximately $4.2\%$ for the discovery reach and $1.3\%$ for the exclusion reach.

For $R_L=0.01$ and $\mathcal{L}=500~\mathrm{fb^{-1}}$, the exclusion reach also remains below $\Gamma_B/M_B=10\%$ over the full mass interval, reaching approximately $6.0\%$ at $M_B=4.5~\mathrm{TeV}$ for $\delta_B=0$. In contrast, the discovery reach approaches the broad-width regime towards the upper end of the mass range. At $M_B=4.5~\mathrm{TeV}$, the corresponding width-to-mass ratio is approximately $19\%$ for $\delta_B=0$ and exceeds the $\Gamma_B/M_B=20\%$ contour when a $10\%$ background systematic uncertainty is included. Therefore, while the high-luminosity exclusion sensitivity remains predominantly within a relatively narrow- or moderate-width regime, the highest-mass discovery region for $R_L=0.01$ should be interpreted with appropriate caution.

Overall, the strongest sensitivity in the $Z$-initiated analysis is obtained for $R_L=0.05$ and $\mathcal{L}=500~\mathrm{fb^{-1}}$, where effective couplings as small as $g^\ast\simeq0.038$ and $g^\ast\simeq0.021$ can be probed at the $5\sigma$ discovery and $95\%$ C.L. exclusion levels, respectively.

\subsection{Comparison with Existing and Future Collider Searches}

Current direct searches at the LHC constrain vector-like $B$ quarks mainly in the mass range below about $1.5~\mathrm{TeV}$. Pair-production searches by ATLAS and CMS typically exclude masses of approximately $M_B\simeq1.2$--$1.6~\mathrm{TeV}$, depending on the assumed branching fractions and VLQ representation~\cite{ATLAS2024VLQpair,CMS2019VLQpair_hadronic,CMS:2022VLQpair}. For single production, where the sensitivity depends more strongly on the electroweak coupling, the reported exclusion reach is generally around $M_B\lesssim1.0$--$1.4~\mathrm{TeV}$ for the coupling scenarios considered~\cite{ATLAS:2023VLQB_single,CMS2024VLQBsingle}.

Future colliders can extend the accessible mass range. Studies at the LHeC and FCC--eh have reported sensitivities reaching approximately $M_B\sim1.4~\mathrm{TeV}$ and $M_B\sim2.2~\mathrm{TeV}$, respectively, depending on the assumed coupling and luminosity~\cite{Shang2022VLQB,Gong2020SingleVLQB,ZengPanZhang2023}. At a $3~\mathrm{TeV}$ CLIC with $\mathcal{L}=5~\mathrm{ab^{-1}}$, the $B\to bZ$ channel has been found to provide a $5\sigma$ discovery sensitivity of approximately $M_B\simeq1.2$--$1.9~\mathrm{TeV}$, while the $95\%$ C.L. exclusion reach can extend up to about $2.4~\mathrm{TeV}$~\cite{Han2022VLQCLIC}. At a future $100~\mathrm{TeV}$ $pp$ collider, vector-like $B$ quark masses of about $2.5$--$3~\mathrm{TeV}$ can be probed for a representative coupling of $g^\ast\sim0.2$~\cite{Han2023EPJC846}.

In comparison, the photon-induced channel considered here probes the $M_B=2$--$3~\mathrm{TeV}$ region, while the $Z$-initiated channel extends the investigated mass range up to $M_B=4.5~\mathrm{TeV}$. For the benchmark $R_L=0.05$, the strongest photon-induced sensitivity reaches approximately $g^\ast\simeq0.263$ for discovery and $g^\ast\simeq0.162$ for exclusion at $M_B=2~\mathrm{TeV}$ and $\mathcal{L}=1000~\mathrm{fb^{-1}}$. The $Z$-initiated channel provides a substantially stronger coupling reach, with corresponding values of about $g^\ast\simeq0.038$ and $g^\ast\simeq0.021$ at $\mathcal{L}=500~\mathrm{fb^{-1}}$. The two channels therefore provide complementary coverage, with the photon-induced mechanism probing the multi-TeV region and the $Z$-initiated mechanism extending the sensitivity towards both higher masses and smaller effective couplings.

A direct one-to-one comparison among different collider studies should nevertheless be made with caution, since the analyses employ different coupling normalizations, flavour-mixing assumptions, VLQ representations, decay channels, collider energies, luminosities, and event-selection strategies. The quoted values should therefore be regarded primarily as an indication of the accessible mass and coupling regimes rather than as model-independent limits.

\section{Conclusion}

\label{sec:conclusion}

In this work, the single production of the singlet-type vector-like $B$ quark at an FCC--$\mu p$ muon--proton collider with a centre-of-mass energy of $\sqrt{s}=24.5$~TeV has been investigated through both photon- and $Z$-initiated production mechanisms. The analysis focuses on the $B \to Zb$ decay mode, with the leptonic decay of the $Z$ boson and the hadronic decay of the accompanying $W$ boson, leading to the final state $ \ell^+\ell^- bjj$. This final state provides strong discrimination between signal and background through the characteristic resonance structures, angular correlations, and efficient $b$-jet identification. The applied kinematic selections significantly suppress the SM background contributions while they retain a sufficient fraction of the signal events. In brief, these findings show that the adopted analysis strategy provides an effective and balanced event selection for the heavy-mass regime.

In the study, the discovery and exclusion sensitivities were evaluated within the Asimov statistical framework for the benchmark mixing values $R_L=0.05$ and $R_L=0.01$. The effect of a $10\%$ background systematic uncertainty was found to produce only a relatively small modification of the sensitivity contours, while increasing the integrated luminosity considerably improves both the discovery and exclusion reaches. For the photon-induced channel, the investigated mass range is $M_B=2$--$3$~TeV. For $R_L=0.05$ and $\mathcal{L}=1000~\mathrm{fb^{-1}}$, the $5\sigma$ discovery reach is approximately $g^\ast\simeq0.263$--$0.328$, while the corresponding $95\%$ C.L. exclusion sensitivity extends to approximately $g^\ast\simeq0.162$--$0.197$. For the smaller mixing value $R_L=0.01$, larger effective couplings are required, with the corresponding discovery and exclusion sensitivities reaching approximately $g^\ast\simeq0.469$--$0.625$ and $g^\ast\simeq0.310$--$0.390$, respectively.

A stronger sensitivity is obtained for the $Z$-initiated production mechanism, which extends the investigated mass range up to $M_B=4.5$~TeV. For $R_L=0.05$ and $\mathcal{L}=500~\mathrm{fb^{-1}}$, the $5\sigma$ discovery reach varies approximately over $g^\ast\simeq0.038$--$0.056$, while the $95\%$ C.L. exclusion sensitivity reaches approximately $g^\ast\simeq0.021$--$0.032$. This represents the most favourable sensitivity obtained among the production mechanisms considered in the present study. Moreover, both the discovery and exclusion reaches remain below $\Gamma_B/M_B=10\%$ throughout the investigated $M_B=2$--$4.5$~TeV mass interval. For $R_L=0.01$, the corresponding discovery and exclusion sensitivities at $\mathcal{L}=500~\mathrm{fb^{-1}}$ are approximately $g^\ast\simeq0.080$--$0.120$ and $g^\ast\simeq0.045$--$0.067$, respectively. While the exclusion sensitivity remains predominantly within a relatively narrow- or moderate-width regime, the discovery reach approaches the broad-width region towards the upper end of the mass range. The corresponding high-mass discovery results should therefore be interpreted with appropriate caution.

In conclusion, the findings of this study demonstrate that photon-induced and $Z$-initiated single production mechanisms at the FCC--$\mu p$ collider provide complementary opportunities for the investigation of heavy singlet VLQ-$B$ quarks. The photon-induced channel provides sensitivity in the $M_B=2$--$3$~TeV region, while the $Z$-initiated channel significantly extends the accessible parameter space up to $M_B=4.5$~TeV and probes considerably smaller effective coupling values. In particular, the strongest sensitivity is obtained for the $Z$-initiated channel with $R_L=0.05$, where effective couplings as small as $g^\ast\simeq0.038$ and $g^\ast\simeq0.021$ can be probed at the $5\sigma$ discovery and $95\%$ C.L. exclusion levels, respectively, in the most favourable part of the parameter space. These results complement the reach of the LHC and other future collider options in the high-mass regime, indicating that $\mu p$ machines can play an important complementary role in the search for physics beyond the SM. In this sense, future $\mu p$ colliders may become a valuable component of the broader accelerator programme aimed at exploring heavy new states.

\section*{Acknowledgements}

This work was supported by the Scientific Research Projects Coordination Unit of
Zonguldak Bülent Ecevit University under Project No.~2025-22794455-04.

\FloatBarrier

\bibliographystyle{unsrt}
\bibliography{references}

\end{document}